\documentclass[onecolumn,nofootinbib,amsmath,showpacs]{article}
\pdfoutput=1 
\usepackage{jheppub} 
\usepackage[T1]{fontenc}
\usepackage{graphicx}
\usepackage{dcolumn}
\usepackage{bm}
\usepackage{color}
\usepackage{float} 
\newcommand{\hs}{\hspace*{0.5cm}}

\newcommand{\be}{\begin{equation}}
\newcommand{\ee}{\end{equation}}
\newcommand{\bea}{\begin{eqnarray}}
\newcommand{\eea}{\end{eqnarray}}
\newcommand{\ben}{\begin{enumerate}}
\newcommand{\een}{\end{enumerate}}
\newcommand{\bde}{\begin{widetext}}
\newcommand{\ede}{\end{widetext}}

\newcommand{\crn}{\nonumber \\}
\newcommand{\Tr}{\mathrm{Tr}}

\newcommand{\al}{\alpha}
\newcommand{\la}{\lambda}

\newcommand{\ga}{\gamma}

\newcommand{\pa}{\partial}

\newcommand{\fr}{\frac}
\newcommand{\bc}{\begin{center}}
\newcommand{\ec}{\end{center}}
\newcommand{\Ga}{\Gamma}

\newcommand{\La}{\Lambda}

\newcommand{\AdrHEPC}{Phenikaa Institute for Advanced Study and Faculty of Basic Science, Phenikaa University, To Huu, Yen Nghia, Ha Dong, Hanoi 100000, Vietnam}
\begin{document}
\allowdisplaybreaks
\title{Neutrino mass and dark matter from an approximate $B-L$ symmetry} 

\author{Duong Van Loi,} 
\emailAdd{loi.duongvan@phenikaa-uni.edu.vn}
\author[1]{Phung Van Dong,\note{Corresponding author at Phenikaa University, Hanoi 100000, Vietnam}} 
\emailAdd{dong.phungvan@phenikaa-uni.edu.vn} 
\author[2]{Dang Van Soa\note{Leave of absence from Hanoi Metropolitan University, Hanoi 100000, Vietnam}}
\emailAdd{dvsoa@assoc.iop.vast.ac.vn}
\affiliation{\AdrHEPC} 

\date{\today}

\keywords{Global symmetry, Neutrino physics, Cosmology of theories beyond the SM}

\arxivnumber{1911.04902}
 
\abstract{
We argue that neutrino mass and dark matter can arise from an approximate $B-L$ symmetry. This idea can be realized in a minimal setup of the flipped 3-3-1 model, which discriminates lepton families while keeping universal quark families and uses only two scalar triplets in order for symmetry breaking and mass generation. This proposal contains naturally an approximate non-Abelian $B-L$ symmetry which consequently leads to an approximate matter parity. The approximate symmetries produce small neutrino masses in terms of type II and III seesaws and may make dark matter long lived. Additionally, dark matter candidate is either unified with the Higgs doublet by gauge symmetry or acted as an inert multiplet. The Peccei-Quinn symmetry is discussed. The gauge and scalar sectors are exactly diagonalized. The signals of the new physics at colliders are examined.}
 
\maketitle
\flushbottom

\section{Introduction}

The standard model is incomplete, since it cannot address neutrino oscillations. Additionally, since the early universe is a quantum system, we need to account for the cosmological challenges of particle physics, such as matter-antimatter asymmetry, dark matter, and cosmic inflation, which all lie beyond the standard model. On the theoretical side, the standard model cannot explain why there are just three fermion families, strong CP conservation, and electric charge quantization.   

Among various approaches to the new physics, the model based upon the gauge symmetry $SU(3)_C\otimes SU(3)_L\otimes U(1)_X$ (3-3-1) \cite{Pisano:1991ee,Frampton:1992wt,Foot:1992rh,Singer:1980sw,Montero:1992jk,Foot:1994ym}  reveals as one of the strong candidates. 
Indeed, the number of fermion families matches that of colors \cite{Singer:1980sw,Frampton:1992wt} and the quantization of electric charge results from particle content \cite{Pisano:1996ht,Doff:1998we,deSousaPires:1998jc,deSousaPires:1999ca,VanDong:2005ux}, which are all due to anomaly cancelation. The model contains automatic Peccei-Quinn symmetries which solve the strong CP question \cite{Pal:1994ba,Dong:2012bf}. It can provide both small neutrino masses by implementing seesaw mechanisms \cite{Tully:2000kk,Dias:2005yh,Chang:2006aa,Dong:2006mt,Dong:2008sw,Dong:2010gk,Dong:2010zu,Dong:2011vb,Dias:2012xp,Catano:2012kw,Boucenna:2014ela,Boucenna:2014dia,Boucenna:2015zwa,Okada:2015bxa,Pires:2014xsa} and dark matter stability by interpreting global or residual gauge symmetry \cite{Fregolente:2002nx,Hoang:2003vj,Filippi:2005mt,deS.Pires:2007gi,Mizukoshi:2010ky,Alvares:2012qv,Profumo:2013sca,Kelso:2013nwa,daSilva:2014qba,Dong:2013ioa,Dong:2014esa,Dong:2015rka,Dong:2013wca,Dong:2014wsa,Dong:2015yra,Huong:2016ybt,Alves:2016fqe}. The inflation and leptogenesis schemes can be realized in this framework \cite{Huong:2015dwa,Dong:2018aak,Dong:2017ayu,Ferreira:2016uao}.  

An interesting proposal of flipped 3-3-1 model has been made \cite{Fonseca:2016tbn}, in which in contrast with the normal setup, one of lepton families transforms differently from the remaining lepton families, whereas all quark families are identical, under $SU(3)_L$ group, in order to cancel the $[SU(3)_L]^3$ anomaly \cite{Gross:1972pv,Georgi:1972bb,Banks:1976yg,Okubo:1977sc}. This inverse of quark and lepton arrangements in comparing to the normal setup consequently converts questions of quark flavors \cite{Ng:1992st,GomezDumm:1994tz,Long:1999ij,Buras:2012dp,Buras:2013dea,Gauld:2013qja,Buras:2015kwd,Dong:2015dxw} to those associate to lepton flavors \cite{Huong:2019vej}. In addition, the theory realizes only a special fermion content which implies dark matter stability by the matter parity as a residual gauge symmetry \cite{Huong:2019vej}\footnote{At high energy, the gauge completion realizes a 3-3-1-1 symmetry \cite{Dong:2013wca,Dong:2014wsa,Huong:2015dwa,Dong:2015yra,Huong:2016ybt,Alves:2016fqe,Dong:2018aak}, which contains noncommutative $B-L$ symmetry and matter parity.}. Note that the matter parity also suppresses unwanted vacua and interactions in order to make the model manifestly. Therefore, the model is very predictive since the lepton flavor violating processes and dark matter observables are all governed by gauge principles as mediated by the new neutral gauge boson $Z'$. On the other hand, the $[\mathrm{gravity}]^2U(1)_X$ anomaly vanishes, which validates the model up to the Planck scale where the effect of quantum gravity enters \cite{Delbourgo:1972xb,AlvarezGaume:1983ig}. 

As studied in \cite{Fonseca:2016tbn,Huong:2019vej} the flipped 3-3-1 model contains a plenty of scalars, including three triplets and one sextet. Such a scalar sector is complicated, containing several Higgs doublets in the triplets and sextet and leading to dangerous lepton flavor violating Higgs decays \cite{Huong:2019vej,Okada:2016whh}. To cure this problem, we propose a minimal scalar sector that contains only two scalar triplets, one for 3-3-1 symmetry breaking and another one for standard model electroweak symmetry breaking. There is no new Higgs doublet presenting in the model. The scalar sector of the model is now calculable and predictive. 

In spite of a minimal scalar content, we argue that all the fermions including neutrinos get appropriate masses. Indeed, the Peccei-Quinn like symmetries are automatically realized in this framework which ensure the strong CP conservation. There is no axion since it is already eaten by a new gauge boson. All ordinary quarks get a mass since the Peccei-Quinn symmetries are completely broken or violated. In addition, the model naturally recognizes an approximate $B-L$ symmetry. Hence, neutrino mass generation schemes arise from such approximate $B-L$ and Peccei-Quinn symmetries. All the interactions are obtained and we make necessary bounds on the model under the current experimental data.     

The rest of this work is organized as follows. In Sec. \ref{model}, we propose the flipped 3-3-1 model with minimal scalar content. Secs. \ref{fer} and \ref{lep} are devoted to determine fermion masses. In Sec. \ref{PQ}, we give a brief comment on Peccei-Quinn symmetry. In Secs. \ref{sca} and \ref{gau}, we study the mass spectra of the scalar and gauge bosons, respectively. Here, the gauge interactions of fermions are also investigated. In Sec. \ref{dart} we interpret dark matter candidates and obtain their observables. The collider phenomena are given in Sec. \ref{collider}. We summarize our results and conclude this work in Sec. \ref{conl}.                         
 
\section{\label{model}Minimal flipped, approximate $B-L$ and matter parity}

As mentioned, the gauge symmetry is \be SU(3)_C\otimes SU(3)_L\otimes U(1)_X,\ee where the last two factors are a nontrivial extension of the standard model electroweak group, while the first factor is the ordinary color group. Additionally, the electric charge and hypercharge are correspondingly embedded in the 3-3-1 symmetry as 
\be Q=T_3+\frac{1}{\sqrt3} T_8 +X,\hs Y=\frac{1}{\sqrt3} T_8 + X,\label{ECO}\ee 
where $T_n$ ($n=1,2,...,8$) and $X$ are $SU(3)_L$ and $U(1)_X$ generators, respectively. The $T_8$ coefficient is fixed to make exotic fermion spectrum phenomenologically viable \cite{Fonseca:2016tbn,Huong:2019vej}. 

The key observation is that the $[SU(3)_L]^3$ anomaly of a sextet equals seven times that of a triplet, $\mathcal{A}(6)=7\mathcal{A}(3)$ \cite{Fonseca:2016tbn,Huong:2019vej}. This leads to a flipped fermion content and solution for family number, in contrast with the usual 3-3-1 approach, such that
\bea \psi_{1L}&=& \left(  \begin{array}{ccc} 
      \xi^{+} & \fr{1}{\sqrt{2}}\xi^{0} & \fr{1}{\sqrt{2}} \nu_1 \\
      \fr{1}{\sqrt{2}} \xi^{0} & \xi^{-} & \fr{1}{\sqrt{2}} e_1 \\
      \fr{1}{\sqrt{2}} \nu_1 & \fr{1}{\sqrt{2}} e_1 & E_1 \\
   \end{array}\right)_L\sim \left(1,6,-\fr{1}{3}\right),\\
   \psi_{\al L}&=&\left(
   \begin{array}{c}
   \nu_{\al }\\
   e_{\al }\\
   E_{\al}\\
   \end{array}
   \right)_L\sim \left(1,3,-\fr{2}{3}\right),\\
   e_{aR} &\sim& (1,1,-1),\hs E_{aR}\sim (1,1,-1),\\
   Q_{aL}&=&\left(
   \begin{array}{c}
   d_a\\
   -u_a\\
   U_a\\
   \end{array}
   \right)_L\sim \left(3,3^*,\fr{1}{3}\right),\\ 
   u_{aR} &\sim& (3,1,2/3),\hs d_{aR }\sim (3,1,-1/3),\hs U_{aR}\sim (3,1,2/3),\eea where $a=1,2,3$ and $\alpha=2,3$ are family indices. The new fields $E,U$ take the same electric charges as $e$ and $u$, respectively. This fermion content is free from all the anomalies, including the gravitational one.   

The previous studies used one scalar sextet and three scalar triplets \cite{Fonseca:2016tbn,Huong:2019vej}. In this work, only two of them are needed,  
\bea &&\rho=\left(
\begin{array}{c}
\rho^+_1\\
\rho^0_2\\
\rho^0_3\\
\end{array}
\right)\sim (1,3,1/3),\hs \chi=\left(
\begin{array}{c}
\chi^+_1\\
\chi^0_2\\
\chi^0_3\\
\end{array}
\right)\sim (1,3,1/3),\eea which would make the model phenomenologically viable. 
They develop vacuum expectation values (VEVs), such as
\bea &&\langle\rho\rangle=\fr{1}{\sqrt2}\left(
\begin{array}{c}
	0\\
	v\\
	w'\\
\end{array}
\right),\hs \langle\chi\rangle=\fr{1}{\sqrt2}\left(
\begin{array}{c}
	0\\
	v'\\
	w\\
\end{array}
\right).\label{29dd}\eea
The 3-3-1 symmetry is broken down to the standard model due to $w$. The standard model gauge symmetry is broken down to $SU(3)_C\otimes U(1)_Q$ due to $v$. To keep consistency with the standard model, we impose  $v\ll w$. Although $\rho$ and $\chi$ have the same gauge charges, they differ in $B-L$ charge, as shown below. Additionally, $\chi_2$ and $\rho_3$ have nonzero $B-L$ charges which lead to the corresponding VEVs to be small, i.e. $v',w'\ll v$, set by the $B-L$ violating scalar potential. This leads to $v',w'\ll v\ll w$. On the other hand, if one chooses another scalar triplet, i.e. $\eta=(\eta^0_1\ \eta^-_2\ \eta^-_3)^T$, instead of $\rho$, it is not naturally to induce\footnote{by radiative corrections or effective interactions.} a large mass for the top quark, since this mass vanishes at tree-level.    

As investigated in \cite{Huong:2019vej}, $B-L$ neither commutes nor closes algebraically with the 3-3-1 symmetry, which is unlike the standard model. Indeed, since $B-L=-\mathrm{diag}(0,0,0,1,1,2)$ for the sextet $(\xi^+\ \xi^0\ \xi^-\ \nu_1\ e_1\ E_1)_L$ and $B-L=-\mathrm{diag}(1,1,2)$ for the triplet $(\nu_\al\ e_\al\ E_\al)_L$, it obeys $[B-L,T_n]\neq 0$ for $n=4,5,6,7$, $\mathrm{Tr}(B-L)\neq 0$ and that $B-L$ is not related to $X$ and ordinary color charges. The noncommutation implies that $B-L$ is a gauged charge and the nonclosure requires a gauge completion of the model which yields a $SU(3)_C\otimes SU(3)_L\otimes U(1)_X\otimes U(1)_N$ (3-3-1-1) symmetry \cite{Dong:2013wca,Dong:2014wsa,Huong:2015dwa,Dong:2015yra,Huong:2016ybt,Alves:2016fqe,Dong:2018aak}, where $B-L=(2/\sqrt{3})T_8+N$ and they are collected in Tabs. \ref{tb1} and \ref{tb2}.
\begin{table}
\bc
\begin{tabular}{|ccccccccccc|}
\hline
Multiplet & $\psi_{1L}$ & $\psi_{\al L}$ & $Q_{aL}$ & $e_{aR}$ & $E_{aR}$ & $u_{aR}$ & $d_{aR}$ & $U_{aR}$ & $\rho$ & $\chi$\\
\hline
$N$ & $-2/3$ & $-4/3$ & $2/3$ & $-1$ & $-2$ & $1/3$ & $1/3$ & $4/3$ & $-1/3$ & $2/3$ \\
\hline
\end{tabular}
\caption{\label{tb1} $N$-charge of particle multiplets.}
\ec
\end{table} 
\begin{table}
\bc
\begin{tabular}{|ccccccccccc|}
\hline
Field & $\xi^{+}$ & $\xi^{0}$ & $\xi^{-}$ & $E^-$ & $U^{2/3}$ & $X^+$ & $Y^0$ & $\rho^0_3$ & $\chi^+_{1}$ & $\chi^0_2$\\
\hline  
$B-L$ & 0 & 0 & 0 & $-2$ & $4/3$ & $1$ & $1$ & $-1$ & $1$ & 1 
\\
$W_P$ & $-1$ & $-1$ & $-1$ & $-1$ & $-1$ & $-1$ & $-1$ & $-1$ & $-1$ & $-1$ \\
\hline
\end{tabular}
\caption{\label{tb2} Nontrivial $B-L$ charge and matter parity.}
\ec
\end{table} The 3-3-1-1 symmetry is broken down to a residual symmetry at low energy, $W_P=(-1)^{3(B-L)+2s}=(-1)^{2\sqrt{3}T_8+3N+2s}$, called matter parity, which determines dark matter stability and prevents unwanted vacua and interactions. 

In this work, to avoid such a 3-3-1-1 extension, i.e. keeping consistency for the 3-3-1 model, we interpret that $B-L$, thus 3-3-1-1 and matter parity, is an approximate symmetry, by contrast. The approximation is naturally recognized by a minimal scalar content, such that (i) the VEVs associate to odd scalar fields spontaneously break $B-L$ and matter parity and (ii) the Yukawa and scalar interactions explicitly violate $B-L$ and matter parity. In addition to the condition of $B-L$ violating scalar potential that sets $u',w'$, the approximate symmetry obviously implies $v',w'\ll v$, since otherwise the $B-L$ conservation ensures $u'=w'=0$ \cite{Huong:2019vej}. 

The total Lagrangian consists of \be \mathcal{L}=\mathcal{L}_{\mathrm{kinetic}}+\mathcal{L}_{\mathrm{Yukawa}}-V.\ee 
The first part contains kinetic terms and gauge interactions,
\bea
\mathcal{L}_{\text{kinetic}}&=&\bar{F}i\gamma^\mu D_\mu F+(D^\mu S)^\dagger (D_\mu S)-\fr{1}{4}G_{n\mu \nu}G_n^{\mu \nu}-\fr{1}{4}A_{n \mu \nu}A_n^{\mu \nu}-\fr{1}{4}B_{\mu \nu}B^{\mu \nu}, 
\eea  
where $F$ and $S$ run over fermion and scalar multiplets, respectively. The covariant derivative and field strength tensors take the forms, $D_\mu=\pa_\mu+ig_s t_{n}G_{n\mu}+igT_n A_{n\mu}+ig_X X B_\mu$, $G_{n \mu \nu}=\partial_{\mu} G_{n\nu}-\partial_{\nu}G_{n \mu}-g_sf_{nmk}G_{m\mu}G_{k\nu}$,
 $A_{n \mu \nu}= \partial_{\mu} A_{n\nu}-\partial_{\nu}A_{n \mu}-g f_{nmk}A_{m\mu}A_{k\nu}$, and $B_{\mu \nu}= \partial_{\mu} B_{\nu}-\partial_{\nu}B_{ \mu}$. Here $(g_s,g,g_X)$, $(t_n,T_n,X)$, and $(G_n,A_n,B)$ are coupling constants, generators, and gauge bosons according to the 3-3-1 subgroups, respectively, and $f_{nmk}$ define $SU(3)$ structure constants. 
 
 The Yukawa interactions are given, up to six dimensions, by
\bea \mathcal{L}_{\mathrm{Yukawa}}&=&h_{ab}^{U}\bar{Q}_{aL}\chi^* U_{bR}+h^E_{\al b}\bar{\psi}_{\al L} \chi E_{bR} +h^u_{ab}\bar{Q}_{aL}\rho^{*}u_{bR}+h^e_{\al b}\bar{\psi}_{\al L}\rho e_{bR}\crn
 &&+s^u_{ab}\bar{Q}_{aL}\chi^{*}u_{bR}+s^e_{\al b}\bar{\psi}_{\al L}\chi e_{bR}+s_{ab}^{U}\bar{Q}_{aL}\rho^* U_{bR}+s^E_{\al b}\bar{\psi}_{\al L} \rho E_{bR} \crn
&&+ \fr{h_{11}^{\xi}}{\La}\bar{\psi}^c_{1L}\chi \chi \psi_{1L}+\fr{h_{1b}^{E}}{\La}\bar{\psi}_{1L}\chi \chi E_{bR}+\fr{h_{1b}^{e}}{\La}\bar{\psi}_{1L}\chi \rho e_{bR}+\fr{h^d_{ab}}{\La}\bar{Q}_{aL}\chi \rho d_{bR} \crn 
&&+\fr{s_{1b}^{E}}{\La}\bar{\psi}_{1L}\chi \rho E_{bR}+\fr{s'^E_{1b}}{\La}\bar{\psi}_{1L}\rho \rho E_{bR}+\fr{s_{1b}^{e}}{\La}\bar{\psi}_{1L}\chi \chi e_{bR}+\fr{{\rm s}_{1b}^{e}}{\La}\bar{\psi}_{1L}\rho \rho e_{bR}
\crn
&&+\fr{s_{11}}{\La}\bar{\psi}^c_{1L}\chi \rho \psi_{1L}+ \fr{s'_{11}}{\La}\bar{\psi}^c_{1L}\rho \rho \psi_{1L}+\frac{s_{1\al}}{\La^2}\bar{\psi}^c_{1L}\chi \chi\rho \psi_{\al L}+\frac{s'_{1\al}}{\La^2}\bar{\psi}^c_{1L}\chi \rho\rho \psi_{\al L} \crn
&&+H.c.,\label{Yukawa} \eea where the couplings $h$ conserve $B-L$ while the couplings $s$ (or $\rm s$) and $s'$ violate $B-L$ by one and two units, respectively. $\La$ is a new physics or cutoff scale that defines (induces) the effective interactions. The Lagrangian might include other six-dimensional operators that directly contribute to the existing renormalizable interactions of the same fermion types, e.g. $\bar{Q}_{aL}\chi^*\chi\rho^{*}u_{bR}$, $\bar{\psi}_{\al L}\chi^* \chi\rho e_{bR}$, and $\bar{\psi}_{\al L}\chi^* \chi \rho E_{bR}$. However, concerning mass generation such contributions are radically smaller than the renormalizable ones, which should be neglected. It is clear that the first and second rows includes renormalizable interactions, giving leading tree-level masses for heavy quarks and leptons. The remainders are nonrenormalizable interactions, providing subleading masses for lighter particles. The interactions that violate $B-L$ and/or matter parity must satisfy the condition $s,s'\ll h$, respectively. 
 
The scalar potential takes the form, 
\bea V&=& \mu_1^2\rho^{\dagger}\rho+\mu_2^2 \chi^{\dagger}\chi+\la_1(\rho^{\dagger}\rho)^2+\la_2(\chi^{\dagger}\chi)^2+\la_3(\rho^{\dagger}\rho)(\chi^{\dagger}\chi)+\la_4(\rho^{\dagger}\chi)(\chi^{\dagger}\rho)\crn
&&+\left[\bar{\mu}^2_3\chi^\dagger \rho +\bar{\la}_5(\chi^\dagger \rho)^2+(\bar{\la}_6\rho^\dagger \rho+\bar{\la}_7\chi^\dagger \chi)\chi^\dagger \rho+H.c.\right].\label{scd}\eea 
Like the Yukawa interactions, the potential parameters $\bar{\la}$ and $\bar{\mu}_3$ violate $B-L$ by one or two units. These violating parameters should be small, i.e. $\bar{\la}\ll \la$ and $\bar{\mu}_3\ll \mu_{1,2}$, since $B-L$ is approximate and the fact that the $B-L$ conservation implies $\bar{\la}=0=\bar{\mu}_3$.

\section{\label{fer}Quark mass}
Quarks gain a mass from the Lagrangian,
\bea \mathcal{L}_{\mathrm{Yukawa}}&\supset&h_{ab}^{U}\bar{Q}_{aL}\chi^* U_{bR}+h^u_{ab}\bar{Q}_{aL}\rho^{*}u_{bR}\crn
&&+s^u_{ab}\bar{Q}_{aL}\chi^{*}u_{bR}+s_{ab}^{U}\bar{Q}_{aL}\rho^* U_{bR}\crn
&&+\fr{h^d_{ab}}{\La}\bar{Q}_{aL}\chi \rho d_{bR} +H.c. \eea

Substituting the VEVs, we obtain a mass matrix for down quarks,
\bea [m_d]_{ab}=\fr{h^d_{ab}}{2\La}(wv -w'v')\simeq \fr 1 2 h^d_{ab} v,\label{32dd} \eea where we use the condition $v',w'\ll v\ll w$ and set $\La\simeq w$ without loss of generality. The up quark mass matrix in basis $(u\ U)$, where $u=(u_1\ u_2\ u_3)$ and $U=(U_1\ U_2\ U_3)$, is
\be m_{\mathrm{up}}=\fr{1}{\sqrt{2}} \left(\begin{array}{cc}
h^u v+s^u v' & h^U v'+s^U v\\
-h^u w'-s^u w & -h^U w-s^U w'
\end{array}
\right).\label{33dd}\ee Using the above condition for VEVs and $s^u,s^U\ll h^u,h^U$, the ordinary quarks $u_a$ and exotic quarks $U_a$ are decoupled, leading to \be [m_u]_{ab}\simeq \fr{1}{\sqrt{2}} h^u v,\hs [m_U]_{ab}\simeq - \fr{1}{\sqrt{2}} h^U w.\ee

Hence, the ordinary up and down quarks get appropriate masses proportional to the weak scale $v$, while the exotic quarks are heavy at the 3-3-1 breaking scale $w$. 

\section{\label{lep}Lepton mass}

The Yukawa interactions of leptons can be divided into $\mathcal{L}_{\mathrm{Yukawa}}\supset \mathcal{L}_{l}+\mathcal{L}_{\nu}$, where the first term provides masses for the ordinary and new charged leptons ($e_a, E_a$), while the second term gives masses for neutrinos and lepton triplet ($\nu_a, \xi$), given respectively by 
\bea \mathcal{L}_{l}&=&h^E_{\al b}\bar{\psi}_{\al L} \chi E_{bR} +h^e_{\al b}\bar{\psi}_{\al L}\rho e_{bR}+s^e_{\al b}\bar{\psi}_{\al L}\chi e_{bR}+s^E_{\al b}\bar{\psi}_{\al L} \rho E_{bR} \crn
&&+\fr{h_{1b}^{E}}{\La}\bar{\psi}_{1L}\chi \chi E_{bR}+\fr{h_{1b}^{e}}{\La}\bar{\psi}_{1L}\chi \rho e_{bR} +\fr{s_{1b}^{E}}{\La}\bar{\psi}_{1L}\chi \rho E_{bR}\crn 
&&+\fr{s'^E_{1b}}{\La}\bar{\psi}_{1L}\rho \rho E_{bR}+\fr{s_{1b}^{e}}{\La}\bar{\psi}_{1L}\chi \chi e_{bR}+\fr{{\rm s}_{1b}^{e}}{\La}\bar{\psi}_{1L}\rho \rho e_{bR} +H.c.,\label{nl1}\\
\mathcal{L}_{\nu}&=& \fr{h_{11}^{\xi}}{\La}\bar{\psi}^c_{1L}\chi \chi \psi_{1L}+\fr{s_{11}}{\La}\bar{\psi}^c_{1L}\chi \rho \psi_{1L}+ \fr{s'_{11}}{\La}\bar{\psi}^c_{1L}\rho \rho \psi_{1L}\crn
&&+\frac{s_{1\al}}{\La^2}\bar{\psi}^c_{1L}\chi \chi\rho \psi_{\al L}+\frac{s'_{1\al}}{\La^2}\bar{\psi}^c_{1L}\chi \rho\rho \psi_{\al L} +H.c. \label{nl2}\eea
Note that there slightly mix between $(e_a,E_a)$ and $(\xi^\pm)$, which are strongly suppressed by the conditions $v',w'\ll v\ll w$ and $s,s'\ll h$, as neglected. 

Substituting the VEVs for scalar fields in (\ref{nl1}), we get the mass matrix for the charged leptons which yields
\bea && [m_e]_{\al b}\simeq -\fr{1}{\sqrt{2}} h^e_{\al b} v,\hs  [m_e]_{1 b}\simeq -\fr{1}{2\sqrt{2}} h^e_{1 b} v,\crn
&& [m_E]_{\al b}\simeq -\fr{1}{\sqrt{2}} h^E_{\al b} w,\hs  [m_E]_{1 b}\simeq -\fr{1}{2} h^E_{1 b} w.\eea The ordinary charged leptons have appropriate masses as in the standard model, while the new leptons $E_a$ have large masses in $w$ scale, as expected. 

For the neutrinos and lepton triplet, we derive their mass Lagrangian from (\ref{nl2}) as
\be
\mathcal{L}_{\nu}\supset -m_{\xi^\pm}\bar{\xi}^\pm\xi^\pm -\fr 1 2 \left(\begin{array}{cccc}
  	\bar{\nu}^c_{1L} &
  	\bar{\nu}^c_{2L}&
  	\bar{\nu}^c_{3L} &
  	\bar{\xi}^{0c} 
  \end{array}\right) M \left(\begin{array}{cccc}
  	\nu_{1L} &
  	\nu_{2L}&
  	\nu_{3L} &
  	\xi^0 
  \end{array}\right)^T+H.c.,
\ee
where $m_{\xi^\pm}\simeq -h^\xi_{11}w$ and $M$ is given by  
\be
M\simeq \left(\begin{array}{cccc}
  	\frac{(s_{11}v^\prime+s^\prime_{11}v)v+h^\xi_{11}v^{\prime 2}}{w} &
-\frac{v}{4w}(s_{12}v^\prime+s^\prime_{12}v)&-\frac{v}{4w}(s_{13}v^\prime+s^\prime_{13}v) & -\fr 1 2 s_{11}v-h^\xi_{11}v' 
 \\
-\frac{v}{4w}(s_{12}v^\prime+s^\prime_{12}v)&0&0&\fr 1 4 s_{12} v\\
-\frac{v}{4w}(s_{13}v^\prime+s^\prime_{13}v)&0&0&\fr 1 4 s_{13} v\\
-\fr 1 2 s_{11}v-h^\xi_{11}v' &\fr 1 4 s_{12} v&\fr 1 4 s_{13} v& h^\xi_{11} w
  \end{array}\right).
\ee 

Due to the conditions $v'\ll v\ll w$ and $s,s'\ll h$, the new neutral lepton $\xi^0$ is heavy and decoupled with a mass, $m_{\xi^0}\simeq h^\xi_{11}w$. Hence, the components of the lepton triplet $\xi$ gain nearly degenerate masses at $w$ scale, $|m_{\xi^0}|\simeq |m_{\xi^\pm}|\sim w$, set by the first term of $\mathcal{L}_\nu $ when $\xi$ interacts with $\chi_3$. 

The neutrinos obtain masses via a combination of type II and III seesaw mechanisms. Indeed, the type III seesaw comes from the interactions of heavy triplet $(\xi^+\ \xi^0\ \xi^-)$ with $(\nu_a \ e_a)(\rho^+_1\ \rho^0_2)$ and with $(\nu_1 \ e_1)(\chi^+_1\ \chi^0_2)$, contained in the Yukawa couplings $s_{1a}$ and $h^\xi_{11}$ in $\mathcal{L}_\nu$, respectively. Whist the type II seesaw arises from the interactions of two lepton doublets with tensor products, $(\rho^+_1\ \rho^0_2)^2$, $(\rho^+_1\ \rho^0_2)(\chi^+_1\ \chi^0_2)$, and $(\chi^+_1\ \chi^0_2)^2$, governed by the Yukawa couplings $s_{1a},\ s'_{1a}$, and $h^\xi_{11}$ in $\mathcal{L}_\nu$. It is clear that such two mechanisms generate appropriate mixing angles between $\nu_1$ and $\nu_{2,3}$ in order to recover the neutrino data, which did not happen in the original model at renormalizable level \cite{Huong:2019vej}. 

By virtue of the seesaw approximation, we have     
\be m_\nu\simeq \fr{v^2}{4h_{11}^\xi  w} \left(\begin{array}{ccc}
  	4h_{11}^\xi s_{11}^\prime-s_{11}^2&\fr 1 2 s_{11}s_{12}-h_{11}^\xi s_{12}^\prime&\fr 1 2 s_{11}s_{13}-h_{11}^\xi s_{13}^\prime\\
  	\fr 1 2 s_{11}s_{12}-h_{11}^\xi s_{12}^\prime&-\fr 1 4 s_{12}^2&-\fr 1 4 s_{12}s_{13}\\
  	\fr 1 2 s_{11}s_{13}-h_{11}^\xi s_{13}^\prime &-\fr 1 4 s_{12}s_{13}&-\fr 1 4 s_{13}^2
  \end{array}\right).\label{nmd}
\ee If $B-L$ is conserved, by contrast, we have $s=s'=0$ (in this case, $v'=w'=0$ as proved in the next section), thus $m_\nu=0$. Hence, the nonzero masses of neutrinos indeed measure an approximate $B-L$ symmetry, governed solely by the $B-L$ violating couplings as seen in~(\ref{nmd}). That said, the neutrino mass matrix is generically small and able to fit the data. 

In addition, the neutrino masses are typically proportional to 
\be m_\nu \sim (s^2 /h,s')\fr{v^2}{w}.\ee Taking $v=246$ GeV, $w\sim 10$ TeV and $m_\nu\sim 0.1$ eV, we derive the $B-L$ violating couplings to be $s'\sim 10^{-10}$ and $s\sim 10^{-5}\sqrt{h}\sim 10^{-6}$, given that $m_\xi\sim 1$ TeV. Thus, the violating couplings $(s,s')$ are smaller than all $B-L$ conserving couplings of charged leptons and quarks, for instance the smallest one $h^e\sim 10^{-5}$ associate to electron and $h^\mu \sim 10^{-3}$ and $h^\tau\sim 10^{-2}$ for muon and tau, respectively. To conclude, the small neutrino masses are suitably generated by an approximate $B-L$ symmetry, characterized by the strength $\epsilon =(s,s')/h\ll 1$, as expected. To make sure of this conclusion, the nature size and stability of approximate symmetry violating strengths responsible for the seesaws and neutrino masses are further determined in the last appendix of this paper. 

It is noted that the triplet lepton $\xi^0$ is $W_P$-odd, playing the main role for neutrino mass generation. However, as shown below, it would fast decay and does not contribute to dark matter, in contrast to~\cite{Huong:2019vej}.

\section{\label{PQ}PQ symmetry}

We now discuss the existence of Peccei-Quinn symmetry in the minimal flipped 3-3-1 model, which subsequently solve the strong CP problem. Let us recall the reader's attention to the original proposal \cite{Peccei:1977hh,Peccei:1977ur} and in kind of the 3-3-1 model \cite{Pal:1994ba,Dong:2012bf}. 

In addition to the 3-3-1 group, we introduce a global Abelian group, called $U(1)_H$. In order for $U(1)_H$ to play the role of Peccei-Quinn symmetry, we require i) the renormalizable Lagrangian is invariant under $U(1)_H$ and ii) the color anomaly $[SU(3)_C]^2 U(1)_H$ does not vanish.  Let $U(1)_H$ charges be $H_Q$, $H_U$, $H_u$, $H_d$, $H_{\psi_1}$, $H_{\psi_\al}$, $H_E$, $H_e$, $H_\chi$, and $H_\rho$ corresponding to the matter multiplets, $Q_{aL}$, $U_{aR}$, $u_{aR}$, $d_{aR}$, $\psi_{1L}$, $\psi_{\al L}$, $E_{aR}$, $e_{aR}$, $\chi$, and $\rho$, respectively. Here note that the repeated flavors should have the same $H$ charge due to the invariance of renormalizable Yukawa Lagrangian. 

Applying the first condition to the scalar potential in (\ref{scd}) and to the first and second lines of the Yukawa Lagrangian in (\ref{Yukawa}), we obtain the following relations,
\bea
&& H_\chi=H_\rho,\\
&& H_Q+H_\chi-H_U=0,\\ 
&& H_{\psi_\alpha}-H_\chi-H_E=0,\\
&& H_Q+H_\chi-H_u=0,\\ 
&& H_{\psi_\alpha}-H_\chi-H_e=0,\\
&& H_Q+H_\rho-H_u=0,\\ 
&& H_{\psi_\alpha}-H_\rho-H_e=0,\\
&& H_Q+H_\rho-H_U=0,\\ 
&& H_{\psi_\alpha}-H_\rho-H_E=0,
\eea
which are equivalently given by 
\bea
&&H_\rho=H_\chi\\
&&H_Q+H_\chi=H_U=H_u,\\
&&H_{\psi_\alpha}-H_\chi=H_E=H_e.
\eea

The second condition implies 
\bea
[SU(3)_C]^2U(1)_H&\sim& \sum_{\text{quarks}}(H_{q_L}-H_{q_R})=3\times 3H_Q-3H_U-3H_u-3H_d\crn
&\sim& 3H_Q-2H_u-H_d=-3H_\chi \neq 0,\label{condition2}
\eea provided that $H_u=H_d$ to simplify the problem. 

Observe that we have 6 equations with 9 variables (except for $H_{\psi_1}$ that is arbitrary). Thus, there is an infinite number of different solutions that satisfy the above conditions. Correspondingly, there is an infinite number of Peccei-Quinn symmetries. Note that this property is actually valid for every 3-3-1 model \cite{Dong:2012bf}. We list, for instance, two different solutions ($H_Q$, $H_U$, $H_u$, $H_d$, $H_{\psi_\al}$, $H_E$, $H_e$, $H_\chi$, $H_\rho$)=$(0,1,1,1,0,-1,-1,1,1)$ or $(0,1,1,1,2,1,1,1,1)$ which are independently linear, where $H_{\psi_1}$ is left arbitrary.   

Since the VEVs $\langle \chi,\rho\rangle$ are simultaneously charged under $SU(3)_L\otimes U(1)_X\otimes U(1)_H$, a residual Peccei-Quinn charge that conserves the vacuum must take the form, \be PQ=a T_3+b T_8+ c X + d H,\ee in which $d\neq 0$. The vacuum annihilation conditions $PQ\langle \chi\rangle =0$ and $PQ\langle \rho\rangle =0$ lead to 
\be
-\frac{a}{2}+\frac{b}{2\sqrt3}+\frac{c}{3}+d H_\chi=0,\hs -\frac{b}{\sqrt3}+\frac{c}{3}+d H_\chi=0.\label{515d}
\ee
We deduce 
\bea PQ= Q+ \delta (H-3H_\chi X),
\eea after substituting the solution $(a,b,c,d)$ and rescaling the $U(1)$ charge by $a$ where $\delta=d/a$. Obviously the Peccei-Quinn symmetry is present after gauge symmetry breaking. This coincides with the fact that the three down quarks have vanishing masses in the renormalizable theory. The strong CP question is solved and there is no axion. Also, there is no Majoron although $B-L$ is broken. As we will see in the following sections, all the Goldstone bosons are eaten by the corresponding gauge bosons.  

As shown in the quark mass section, the down quarks can get consistent masses via the effective interaction, $\bar{Q}_{aL}\chi\rho d_{bR}$, which explicitly violates $U(1)_H$, since $-H_Q+H_\chi+H_\rho+H_d=3H_\chi\neq 0$. It is noteworthy that the neutrino masses can also come from the violation of the Peccei-Quinn symmetry, since $\mathcal{L}_\nu$ in (\ref{nl2}) is generally not invariant under $U(1)_H$ for a generic $H_{\psi_1}$ charge.   

\section{\label{sca}Scalar sector}

Let us expand the scalar fields around their VEVs,
\be 
\rho=\left(
\begin{array}{c}
	\rho_1^+\\
	\fr{v+S_2+iA_2}{\sqrt{2}}\\
	\fr{w^\prime+S_3^\prime+iA_3^\prime}{\sqrt{2}}\\
\end{array}
\right),\hs \chi=\left(
\begin{array}{c}
	\chi^+_1\\
	\fr{v^\prime+S_2^\prime+iA_2^\prime}{\sqrt{2}}\\
	\fr{w+S_3+iA_3}{\sqrt{2}}\\
\end{array}
\right).
\ee
The scalar potential is correspondingly summed of $V = V_{\text{min}}+V_{\text{linear}}+V_{\text{mass}}+V_{\text{int} }$. The first and last terms define vacuum energy and scalar self-interactions, which will be skipped. 

The second term provides conditions of potential minimization,  
\bea 
\mu_1^2 v&+&\la_1 (v^2+w^{\prime 2})v+\la_3\fr{(w^2+v^{\prime 2})v}{2}+\la_4\fr{(w w^\prime+v v^\prime)v^\prime}{2}\crn
&+&\bar{\mu}_3^2 v^\prime+\bar{\lambda}_5(w w^\prime+v v^\prime)v^\prime+\bar{\lambda}_6\frac{2wv w^\prime+3 v^2 v^\prime+ w^{\prime 2}v^\prime}{2}+\bar{\lambda}_7\frac{(w^2+ v^{\prime 2})v^\prime}{2} = 0,\label{ctt1}\eea\bea
\mu_2^2 w&+&\la_2 (w^2+v^{\prime 2})w+\la_3\fr{(v^2+w^{\prime 2})w}{2}+\la_4\fr{(w w^\prime+v v^\prime)w^\prime}{2}\crn
&+&\bar{\mu}_3^2 w^\prime+\bar{\lambda}_5(w w^\prime+v v^\prime)w^\prime
+\bar{\lambda}_6\frac{(v^2+ w^{\prime 2})w^\prime}{2}+\bar{\lambda}_7\frac{3w^2 w^\prime+ 2w v v^\prime+w^\prime v^{\prime 2}}{2} = 0,\label{ctt2}\eea\bea
\mu_2^2 v^\prime &+&\la_2 (w^2+v^{\prime 2})v^\prime+\la_3\fr{(v^2+w^{\prime 2})v^\prime}{2}+\la_4\fr{(w w^\prime+v v^\prime)v}{2}\crn
&+&\bar{\mu}_3^2 v+\bar{\lambda}_5(w w^\prime+v v^\prime)v
+\bar{\lambda}_6\frac{(v^2+ w^{\prime 2})v}{2}+\bar{\lambda}_7\frac{w^2 v+ 2w w^\prime v^\prime+3 v v^{\prime 2}}{2} = 0,\label{ctt3}\eea\bea
\mu_1^2 w^\prime &+&\la_1 (v^2+w^{\prime 2})w^\prime+\la_3\fr{(w^2+v^{\prime 2})w^\prime}{2}+\la_4\fr{(w w^\prime+v v^\prime)w}{2}\crn
&+&\bar{\mu}_3^2 w+\bar{\lambda}_5(w w^\prime+v v^\prime)w+\bar{\lambda}_6\frac{wv^2+3w w^{\prime 2}+2v w^\prime v^\prime}{2}+\bar{\lambda}_7\frac{(w^2+ v^{\prime 2})w}{2} = 0.\label{ctt4}
\eea 
Generally under an approximate $B-L$ symmetry, the violating parameters should be radically smaller than the corresponding conserving ones, $\bar{\mu}\ll \mu$ and $\bar{\la}\ll \la$, since otherwise the $B-L$ conservation demands $\bar{\mu}=0=\bar{\la}$. From (\ref{ctt3}) and (\ref{ctt4}), we derive
\bea &&v'\simeq -\fr{[\bar{\mu}^2_3 + (\bar{\la}_6/2)v^2+(\bar{\la}_7/2)w^2]v}{\mu^2_2+\la_2 w^2+(\la_3/2)v^2+(\la_4/2)(v^2+w v t)}\sim \fr{\bar{\mu}^2_3}{w},\\
&& w'\simeq -\fr{[\bar{\mu}^2_3 + (\bar{\la}_6/2)v^2+(\bar{\la}_7/2)w^2]w}{\mu^2_1+\la_1 v^2+(\la_3/2)w^2+(\la_4/2)(w^2+wv/t)}\sim \fr{\bar{\mu}^2_3}{w},\label{67d}\eea where $t\equiv w'/v'$ is finite and we assume $\bar{\mu}^2_3\sim \bar{\la}_6 v^2\sim \bar{\la}_7 w^2$. Since $\bar{\mu}_3\ll \mu_{1,2}\sim (v,w)$, we get $u',w'\ll \bar{\mu}_3$. In practice, taking $\bar{\mu}_3=1$ MeV to 1 GeV, it implies $u'\sim w'\sim 0.1$ eV to 0.1 MeV, respectively. Such VEVs are responsible for the mentioned neutrino mass generation scheme. Since $u',w'\ll v,w$, from (\ref{ctt1}) and (\ref{ctt2}), we deduce,
\bea v^2\simeq \fr{-\mu^2_1\la_2+\mu^2_2 \la_3/2}{\la_1\la_2-\la^2_3/4},\hs w^2\simeq \fr{-\mu^2_2\la_1+\mu^2_1 \la_3/2}{\la_1\la_2-\la^2_3/4}.\eea Of course, necessary conditions for the potential parameters are \be \la_{1,2}>0,\ \la_3+\la_4\theta(-\la_4)>-2\sqrt{\la_1\la_2},\ \mu^2_1<0,\ \mu^2_2<0,\ee which are required in order for the potential to be bounded from below and to yield the desirable vacuum structure.

The third term in the potential expansion yields scalar mass spectrum. Let us rewrite $V_{\text{mass}}=V_{\text{mass}}^S+V_{\text{mass}}^A$, since all charged scalars have vanishing mass to be identified as the Goldstone bosons of $W$ and $X$ gauge fields, i.e. $G_W^\pm=\rho_1^\pm,\ G_X^\pm=\chi_1^\pm$. The real parts obey a $4\times 4$ mass matrix, \be V^{S}_{\mathrm{mass}} = \frac{1}{2}\left(\begin{array}{cccc}
  	S_2 &
  	S_3&
  	S_2^\prime &
  	S_3^\prime 
  \end{array}\right) M_S^2 \left(\begin{array}{cccc}
  	S_2 &
  	S_3&
  	S_2^\prime &
  	S_3^\prime 
  \end{array}\right)^T,\ee
which provides a Goldstone boson ($G_1$) and a new Higgs field ($\mathcal{H}$) orthogonal to it, 
\be
G_1 \simeq  \frac{w S_2^\prime - vS_3^\prime}{\sqrt{w^2+v^2}},\hs
\mathcal{H}\simeq \frac{v S_2^\prime +w S_3^\prime}{\sqrt{w^2+v^2}},\hs m^2_{\mathcal{H}}\simeq \fr{\la_4}{2}(w^2+v^2).
\ee
There remains a slight mixing between $S_2$ and $S_3$, which yields physical states \be H\simeq S_2-\epsilon S_3,\hs H_1=S_3+\epsilon S_2,\ee where the mixing parameter and masses are given by 
\be \epsilon \simeq \frac{\la_3}{2\la_2}\fr{v}{w},\hs m^2_H\simeq\frac{(4\la_1\la_2-\la_3^2)v^2}{2\la_2},\hs m^2_{H_1}\simeq 2\la_2w^2.
\ee  
The imaginary parts mix via a $4\times 4$ mass matrix, which leads to three Goldstone bosons and a massive pseudo-scalar, 
\bea
G_{Z} &\simeq &  A_2,\hs G_{Z^\prime} \simeq  A_3 ,\hs G_{2}\simeq \frac{w A_2^\prime+v A_3^\prime}{\sqrt{w^2+v^2}},\crn
\mathcal{A}&\simeq &\frac{-vA_2^\prime+wA_3^\prime}{\sqrt{w^2+v^2}},\hs m^2_{\mathcal{A}}\simeq \frac{\la_4}{2}(w^2+v^2).
\eea 

Note that $H$ is light, identical to the standard model Higgs boson. Additionally, we can define $H'=(\mathcal{H}+i\mathcal{A})/\sqrt{2}$ since $\mathcal{H}$ and $\mathcal{A}$ have the same mass, similarly for $G^0_{Y}=(G_1+iG_2)/\sqrt{2}$ to be the Goldstone boson of $Y$ gauge field. In the effective limit, $v\ll w$, the Higgs spectrum is 
\be \rho\simeq \left(
\begin{array}{c}
	G^+_W\\
	\fr{1}{\sqrt{2}}(v+H+i G_Z)\\
	\fr{1}{\sqrt{2}}w^\prime+H'\\
\end{array}
\right),\hs \chi\simeq \left(
\begin{array}{c}
	G^+_X\\
	\fr{1}{\sqrt{2}}v^\prime+G^0_Y\\
	\fr{1}{\sqrt{2}}(w+H_1+iG_{Z'})\\
\end{array}
\right).\label{615d}
\ee 

Apart from the standard model Higgs boson, there are two new Higgs fields: $H_1$ is $W_P$-even, responsible for the 3-3-1 breaking, while $H'$ is $W_P$-odd, unified with the Higgs doublet by the 3-3-1 symmetry \cite{Huong:2019vej}. Even if $H'$ is the lightest $W_P$-particle, it may decay due to the approximate matter parity. Also due to this approximate nature, the candidate may reveal a lifetime longer than our universe's age responsible for dark matter (see below).  

For completion, we supply Appendix \ref{appa} to prove that the VEVs as given obey stable minimum. Additionally, we include Appendix \ref{appb} to present the nature of the VEV alignments under the gauge symmetry. Last, it is noteworthy that the small values of $v',w'$ as well as the hierarchies $v',w'\ll v$ can always be maintained against radiative corrections, as shown in Appendix \ref{appc}.  
     
\section{\label{gau}Gauge sector}

The presence of $v',w'$ leads to a small mixing between the charged gauge bosons, 
\be W^\pm=(A_1\mp iA_2)/\sqrt{2},\hs X^\pm=(A_4\mp iA_5)/\sqrt{2},\ee as well as between the neutral gauge bosons $A_3,A_8,B$ with $A_6$---the real part of non-Hermitian gauge boson, 
\be Y^{0,0*}=(A_6\mp i A_7)/\sqrt{2}.\ee 

However, as proved, $v'$ and $w'$ are strongly suppressed, hence such mixings can be neglected. This leads to the physical gauge bosons $W,\ X$ and $Y$ by themselves with corresponding masses, obtained by
\be m^2_W\simeq \fr{g^2v^2}{4},\hs m^2_X\simeq \fr{g^2w^2}{4},\hs m^2_Y\simeq \fr{g^2(v^2+w^2)}{4}.\ee
Here $W$ is identical to the standard model weak boson, implying $v=246$ GeV. The neutral gauge boson $Y$ is $W_P$-odd and may be long-lived. However, it does not contribute to dark matter, since it annihilates completely, before freeze-out, into $W$ bosons via gauge self-interactions (cf. \cite{Dong:2013wca}). 

Similarly, the neutral gauge bosons are identified by 
\bea 
A &=& s_W A_{3}+c_W\left(\fr{t_W}{\sqrt{3}}A_{8}+\sqrt{1-\fr{t_W^2}{3}}B\right), \\
Z &=&c_W A_{3}-s_W\left(\fr{t_W}{\sqrt{3}}A_{8}+\sqrt{1-\fr{t_W^2}{3}}B\right), \\
 Z'&=&\sqrt{1-\fr{t_W^2}{3}}A_{8}-\fr{t_W}{\sqrt{3}}B,
\eea with the corresponding masses
\be m_A=0,\hs m^2_Z\simeq \fr{g^2v^2}{4c_W^2},\hs m^2_{Z'}\simeq \fr{g^2[c^2_{2W}v^2+4c^4_W w^2]}{4c^2_W(3-4s^2_W)},\ee where
the sine of the Weinberg angle is defined by $s_W = \sqrt{3}t_X/\sqrt{3+4t^2_X}$ with $t_X=g_X/g$. 

The photon has zero mass and decoupled as a physical field, while there is a small mixing between $Z$ and $Z'$, determined by $m^2_{ZZ'}\simeq -g^2c_{2W}v^2/(4c^2_W\sqrt{3-4s^2_W})$. Therefore, this defines the $Z$-$Z'$ mixing angle,  
\be t_{2\varphi}\simeq -\fr{c_{2W}\sqrt{1+2c_{2W}}v^2}{2c_W^4w^2},\ee and the physical neutral fields $Z_1=c_\varphi Z-s_\varphi Z'$ and $Z_2=s_\varphi Z+c_\varphi Z'$. Additionally, this shift in $Z$ mass modifies the $\rho$ parameter to be  
\be \Delta \rho = \fr{m^2_W}{c^2_W m^2_{Z_1}}-1\simeq \fr{(1-t^2_W)^2v^2}{4w^2}.\ee From the global fit $\Delta \rho <0.00058$ \cite{Tanabashi:2018oca}, we deduce $w>3.6$~TeV. This implies $m_{Z'}>1.43$~TeV.  

Using the above results, we have the interactions of $Z$ and $Z'$ with fermions. The vector and axial-vector couplings are listed in Tables \ref{coupz} and \ref{coupzp}.
\allowdisplaybreaks
\begin{table}[!h]
	\bc
	\begin{tabular}{|c|c|c|}
		\hline
		$f$ & $g^{Z}_V(f)$ & $g^{Z}_A(f)$ \\
		\hline 
		$\nu_a $ & $\fr 1 2$ & $\fr 1 2$ \\ 
		\hline
		$e_a$ & $-\fr 1 2 +2s^2_W$ & $-\fr 1 2$ \\ 
		\hline
		$u_a$ &$\fr 1 2 -\fr 4 3 s^2_W$ & $\fr 1 2$\\
		\hline
		$d_a$ &$-\fr 1 2+\fr 2 3 s^2_W$ &  $-\fr 1 2$\\
		\hline
		$\xi^+$ & $2c^2_W$ &  $0$ \\
		\hline
		$E_a$ & $2s^2_W$ & $0$\\
		\hline
		$U_a$ &$-\frac{4}{3}s^2_W$ &  $0$\\
		\hline 
	\end{tabular}
	\caption{\label{coupz} The couplings of $Z$ with fermions.}
	\ec
\end{table}
\begin{table}[!h]
	\bc
	\begin{tabular}{|c|c|c|}
		\hline
		$f$ & $g^{Z'}_V(f)$ & $g^{Z'}_A(f)$ \\
		\hline 
		$\nu_1 $ & $-\frac{c_{2W}}{2\sqrt{1+2c_{2W}}}$ & $-\frac{c_{2W}}{2\sqrt{1+2c_{2W}}}$ \\ 
		\hline 
		$\nu_\al $ & $\frac{1}{2\sqrt{1+2c_{2W}}}$ & $\frac{1}{2\sqrt{1+2c_{2W}}}$ \\  
		\hline
			$e_1$ & $\frac{1-2c_{2W}}{2\sqrt{1+2c_{2W}}}$ & $-\frac{1}{2\sqrt{1+2c_{2W}}}$ \\ 
		\hline
			$e_\al$ & $\frac{2-c_{2W}}{2\sqrt{1+2c_{2W}}}$ & $\frac{c_{2W}}{2\sqrt{1+2c_{2W}}}$ \\ 
		\hline
		$u_a$ &$\frac{c_{2W}-4}{6\sqrt{1+2c_{2W}}}$ & $-\frac{c_{2W}}{2\sqrt{1+2c_{2W}}}$\\
		\hline
		$d_a$ &$-\frac{\sqrt{1+2c_{2W}}}{6}$ &  $-\frac{1}{2\sqrt{1+2c_{2W}}}$\\
		\hline
		$\xi^+$ & $0$ &  $\frac{2c_{W}^2}{\sqrt{1+2c_{2W}}}$ \\
		\hline
		$\xi^0$ & $\frac{c_{W}^2}{\sqrt{1+2c_{2W}}}$ &  $\frac{c_{W}^2}{\sqrt{1+2c_{2W}}}$ \\
		\hline
		$E_1$ & $-\frac{2c_{2W}}{\sqrt{1+2c_{2W}}}$ & $-\frac{2c_{W}^2}{\sqrt{1+2c_{2W}}}$\\
		\hline 
		$E_\al$ & $\frac{1-3c_{2W}}{2\sqrt{1+2c_{2W}}}$ & $-\frac{c_{W}^2}{\sqrt{1+2c_{2W}}}$\\
		\hline
		$U_a$ &$\frac{7c_{2W}-1}{6\sqrt{1+2c_{2W}}}$ &  $\frac{c_{W}^2}{\sqrt{1+2c_{2W}}}$\\
		\hline 
	\end{tabular}
	\caption{\label{coupzp} The couplings of $Z'$ with fermions.}
	\ec
\end{table}

\section{\label{dart}Dark matter} 

As investigated in \cite{Huong:2019vej}, if the $B-L$ symmetry was exact, it would be broken down to a matter parity as the residual gauge symmetry, which subsequently stabilized the dark matter candidates, $H'$ and $\xi^0$. 

In the model under consideration $B-L$, thus resultant matter parity, is an approximate symmetry, implying dark matter instability. However, we expect that the approximate symmetry would suppress $H', \xi^0$ decays or provide other candidates. Let us evaluate their lifetimes, in order to interpret a possible candidate for dark matter.

The scalar candidate interacts with the standard model Higgs boson via 
\be \mathcal{L}\supset -\fr{1}{2}(2\la_1 w'+\bar{\la}_6w)H^2\Re(H'),\ee where $\Re(H')=\mathcal{H}$, while $\Im(H')=\mathcal{A}$ only appears in pair in interactions, including the Yukawa and gauge interactions. Hence, $\mathcal{A}$ is stabilized. For the real part, we obtain the decay rate, $\Ga(\mathcal{H}\rightarrow HH)\simeq |2\la_1 w'+\bar{\la}_6 w|^2/(32\pi m_{\mathcal{H}})$, which leads to a lifetime,
\bea \tau_{\mathcal{H}}&\simeq& \fr{1}{\Ga(\mathcal{H}\rightarrow HH)}\crn
&\simeq& \left(\fr{m_{\mathcal{H}}}{1\ \mathrm{TeV}}\right)\times \left(\fr{10^{-18}}{\bar{\la}_6+2\la_1(w'/w)}\right)^2\times 10\ \mathrm{Gyr},\eea provided that $w=10$ TeV and $\la_1 w'\sim \bar{\la}_6 w$.

The fermion candidate interacts with the standard model lepton and Higgs boson via
\be \mathcal{L}\supset s_{1a}(\xi^+\ \xi^0\ \xi^-)(\nu_{a}\ e_{a})(\rho^+_1\ \rho^0_2)\supset s_{1a}\xi^0\nu_{a}H,\ee which leads to $\xi^0\rightarrow \nu_a H$ decay, yielding 
\be \tau_{\xi^0}\sim \left(\fr{1\ \mathrm{TeV}}{m_{\xi^0}}\right)^3\times \left(\fr{10^{-18}}{s_{1a}}\right)^2\times 10\ \mathrm{Gyr}.\ee

Remarks are in order
\ben
\item The neutrino mass generation requires $s_{1a}\sim 10^{-6}$. Thus, $\xi^0$ fast decays, not contributing to dark matter. 
\item Note that $\bar{\la}_6$ and $w'$ are relaxed from the neutrino mass constraint. We require $\bar{\la}_6\sim w'/w\sim 10^{-18}$, such that $\mathcal{H}$ has a lifetime comparable to the universe's age. In fact, the current bound of dark matter lifetime is a billion times longer than the universe's age \cite{Ackermann:2012qk}, which leads to $\bar{\la}_6\sim w'/w\sim 10^{-23}$. Such tiny values should be strongly suppressed by the approximate $B-L$ symmetry, which is unlike that from the neutrino mass bound. An extra analysis supplied in Appendix \ref{appd} show that the stability of $\bar{\la}_6$ under radiative corrections may only be kept at the violating strength level for generating neutrino mass, i.e. $\bar{\la}_6\lesssim 10^{-10}$, which subsequently rules out the $\mathcal{H}$ candidate.      
\item The pseudo-scalar $\mathcal{A}$ is a natural candidate for dark matter. It is a standard model singlet, hence interacting with ordinary particles only via the Higgs, $H$ and $H_1$, portals. In the early universe, they annihilate to the standard model particles via $s$-channels that set the density, 
\be \Omega h^2\simeq \fr{0.1\ \mathrm{pb}}{\langle \sigma v\rangle}\simeq 0.1  \left(\fr{m_{\mathcal{A}}}{\la'\times 2.7\ \mathrm{TeV}}\right)^2, \ee where $\la'\equiv \la_1 + \la_3(\la_3+\la_4)/[4(\la_4-\la_2)]$. Comparing to the data, $\Omega h^2\simeq 0.1$ \cite{Tanabashi:2018oca}, we obtain $m_{\mathcal{A}}=\la'\times 2.7\ \mathrm{TeV}\sim \mathcal{O}(1)$ TeV. The dark matter $\mathcal{A}$ can scatter off ordinary quarks via $t$-channel $H$-exchange that sets its direct detection cross-section, \be \sigma_{\mathcal{A}-p,n}\simeq \left(\fr{\la_1}{0.1}\right)^2\left(\fr{1\ \mathrm{TeV}}{m_{\mathcal{A}}}\right)^2\times 2.45\times 10^{-45}\ \mathrm{cm}^2.\ee Note that $\la_1$ is proportional to the standard model Higgs coupling. When $\mathcal{A}$ has a correct density, the model predicts $\sigma_{\mathcal{A}-p,n}\sim 10^{-45}\ \mathrm{cm}^2$, coinciding with the direct detection experiment \cite{Aprile:2017iyp} for the dark matter mass in TeV regime.      
\item Following a completely different approach \cite{Dong:2013ioa,Dong:2014esa,Dong:2015dxw}, the minimal flipped 3-3-1 model with approximate $B-L$ symmetry can have naturally a room for extra inert scalars that are odd under a $Z_2$ symmetry. Indeed, the inert scalar multiplets may be a hidden triplet $\eta=(\eta^0_1\ \eta^-_2\ \eta^-_3)$ or hidden sextet $S=(S^{++}_{11}\ S^+_{12}\ S^+_{13}\ S^0_{22}\ S^0_{23}\ S^0_{33})$, which were presented in the complete version \cite{Huong:2019vej}, but now $Z_2$ odd. The dark matter candidate is the lightest inert particle, which may be a singlet or doublet or triplet scalar as resided in $\eta,S$. The phenomenology of dark matter is analogous to \cite{Dong:2013ioa,Dong:2014esa,Dong:2015dxw}, which will be skipped as out of the scope of this work.
\een

\section{\label{collider}Collider search} 

The LEPII experiment searches for new neutral gauge bosons $Z^\prime$ through the channel $e^+e^-\rightarrow f\bar{f}$, where $f$ is some fermion, e.g. $f=\mu$. Because the collision energy is much smaller than the $Z'$ mass, we can integrate $Z'$ out of the Lagrangian and obtain corresponding effective interactions, 
\bea
\mathcal{L}_{\text{eff}}&=&\frac{g^2c_{2W}}{4c_W^2(1+2c_{2W})m^2_{Z^\prime}}(\bar{e}\gamma^\mu P_Le)(\bar{\mu}\gamma^\mu P_L \mu)+(LR)+(RL)+(RR),
\eea where the last terms differ from the first one by chiral structures. Such chiral couplings were extensively studied \cite{Tanabashi:2018oca}. Let us take a typical bound,
\bea
\frac{g^2c_{2W}}{4c_W^2(1+2c_{2W})m^2_{Z^\prime}}<\frac{1}{(6\text{ TeV})^2},
\eea which translates to $m_{Z^\prime}>1.13$ TeV. 

Our previous study for the LHC dilepton and dijet signals in the general flipped 3-3-1 model implied a bound $m_Z'>2.8$ TeV \cite{Huong:2019vej}. Such bound can be applied for the considering model, since in the effective limit ($w\gg v$) the $Z'$ couplings with leptons and quarks are identical to the previous version. 

The new Higgs $H_1$ can be produced at LHC by gluon fusion and then decays to diphoton and/or diboson signals. A naive estimation shows that the diboson signals are negligible, while the remaining productions are $\sigma(pp\to Z'\to \gamma\gamma)\sim \sigma(pp\to Z'\to \gamma Z)\sim \la_2$ fb, which are below the current bound $\sim 1$ fb \cite{Tanabashi:2018oca} for $\la_2$ below the perturbative limit. 

With the dark matter masses ($\mathcal{A}$, $\mathcal{H}$) in TeV regime, they can be created at the LHC due to exchanges of $H_1$ and $Z'$ via couplings \be \mathcal{L}\supset -\fr{(\la_3+\la_4)w}{2} H_1\mathcal{A}\mathcal{A}-\fr{g}{\sqrt{3-t^2_W}}Z'\mathcal{H}\stackrel{\leftrightarrow}{\pa}\mathcal{A}\ee as well as those of $H_1,Z'$ to ordinary particles. The mono-$X$ signature is a jet via $H_1$-exchange $gg\rightarrow g \mathcal{A}\mathcal{A}$, $q\bar{q}\rightarrow g \mathcal{A}\mathcal{A}$, and $gq\rightarrow q \mathcal{A}\mathcal{A}$ as well as via $Z'$ exchange $q\bar{q}\rightarrow g \mathcal{H}\mathcal{A}$ and $gq\rightarrow q \mathcal{H}\mathcal{A}$. Since the mediators $H_1,Z'$ are heavy, such processes can be induced by effective interactions after integrating them out, 
\bea \mathcal{L}_{\mathrm{eff}} &=& -\fr{(\la_3+\la_4)\al_s}{8\pi m^2_{H_1}}G_{n\mu\nu}G_n^{\mu\nu}\mathcal{A}\mathcal{A}\crn
&&-\fr{g^2}{2\sqrt{3-4s^2_W}m^2_{Z'}}\bar{q}\ga_\mu [g^{Z'}_V(q)-g^{Z'}_A(q)\ga_5]q\mathcal{H}\stackrel{\leftrightarrow}{\pa}\mathcal{A}. \eea Generalizing the result in \cite{Belyaev:2018pqr}, we obtain the bounds,
\bea &&\fr{|\la_3+\la_4|\al_s}{8\pi m^2_{H_1}} < \left(\fr{1}{3\ \mathrm{TeV}}\right)^2,\\
&& \fr{g^2}{2\sqrt{3-4s^2_W}m^2_{Z'}} g_m< \left(\fr{1}{0.3\ \mathrm{TeV}}\right)^2,\eea where $g_
m=\mathrm{max}\{|g^{Z'}_V(q)|,|g^{Z'}_A(q)|\}=(3+2s^2_W)/[6\sqrt{3-4s^2_W}]$. The first condition leads to $m_{H_1}>200\sqrt{|\la_3+\la_4|}$, which is below 1 TeV due to perturbative limit $\la_{3,4}<4\pi$, while the second condition implies $m_{Z'}>74$ GeV. Indeed, since the dark matter and mediators are in TeV regime, the monojet signals are negligible.        
    
\section{\label{conl}Conclusion}

We have shown that the neutrino masses and dark matter can be addressed in a common framework that includes approximate non-Abelian $B-L$ symmetry. 

Interpreted in the minimal flipped 3-3-1 model, the neutrino masses automatically come from a combination of type II and III seesaws, while the natural smallness of masses are protected by the approximate symmetry.  

The generation of neutrino and quark masses violates the Peccei-Quinn symmetry, while such symmetry preserves strong CP symmetry and has no Axion.

In the present model, dark matter arises from a $B-L$-charged pseudo-scalar ($\mathcal{A}$) which is unified with the Higgs doublet in a gauge triplet. This candidate has the appropriate density and detection cross-sections. The existence possibility of other candidates such as its real part $\mathcal{H}$ and inert scalars have been examined. 

 The new fermions, gauge and Higgs bosons are obtained, having masses at TeV scale. The new physics effects at colliders are discussed, showing the viability of the model.          
 
\acknowledgments

This research is funded by Vietnam National Foundation for Science and Technology Development (NAFOSTED) under grant number 103.01-2016.44.

\appendix

\section{\label{appa} Stable minimum}

For simplicity, as appropriate to the results/conclusions, we assume that the scalar potential and vacuum conserve CP, i.e. $\bar{\mu}^2_3$, $\bar{\la}_{5,6,7}$, and the VEVs $\langle \rho\rangle=(0,v,w')^T/\sqrt{2}$ and $\langle \chi\rangle=(0,v',w)^T/\sqrt{2}$ are all real. For brevity, we define new variables $a=\sqrt{\langle \rho\rangle ^\dagger \langle \rho\rangle }$, $b=\sqrt{\langle \chi\rangle ^\dagger \langle \chi\rangle }$, and $t=\cos(\langle \rho\rangle, \langle \chi\rangle )$---the cosine of the angle between two vectors $\langle \rho\rangle $ and $\langle \chi\rangle $, hence $\langle \chi\rangle ^\dagger \langle \rho\rangle=\langle \rho\rangle ^\dagger \langle \chi\rangle =abt$.\footnote{The notations $a,b,t$ are different from those in the body text, which should be confused.} The corresponding potential value takes the form,
\be V=\mu^2_1 a^2 +\mu^2_2 b^2+\la_1 a^4 +\la_2 b^4 +\la_3 a^2 b^2+\la_{45} a^2b^2 t^2+\mu^2_3 ab t +\la_6 a^3b t+\la_7 a b^3 t,\label{addp}\ee where we also define $\mu^2_3=2\bar{\mu}^2_3$, $\la_{5,6,7}=2\bar{\la}_{5,6,7}$, and $\la_{45}=\la_4+\la_5$. 

The extremum conditions read
\bea \fr{\pa V}{\pa a} &=& 2\mu^2_1 a +4 \la_1 a^3 + 2 \la_3 a b^2+(2 \la_{45} a b t+\mu^2_3  +3 \la_6 a^2+\la_7 b^2 )bt=0,\label{ad1l}\\
\fr{\pa V}{\pa b}&=& 2 \mu^2_2 b + 4 \la_2 b^3 +2\la_3 a^2 b +(2 \la_{45} ab t+\mu^2_3 +\la_6 a^2 +3\la_7 b^2 )at=0,\label{ad2l}\\
\fr{\pa V}{\pa t}&=&(2\la_{45} a b t +\mu^2_3  +\la_6 a^2 +\la_7  b^2)ab =0.\eea
The last equation leads to a solution, $ab=0$ or $2\la_{45} a b t +\mu^2_3  +\la_6 a^2 +\la_7  b^2=0$. However, the first solution $ab=0$ implies $a=b=0$ with the aid of (\ref{ad1l},\ref{ad2l}) and the potential parameter conditions in the body text. This defines a local maximum, i.e. vacuum instability, which should be neglected. We consider the second solution, which obeys
\be t=-\fr{\mu^2_3+\la_6a^2+\la_7b^2}{2\la_{45}ab}.\label{ad3l}\ee Substituting $t$ to (\ref{ad1l},\ref{ad2l}) yields
\bea &&(4\la_1-\la^2_6/\la_{45})a^2+(2\la_3-\la_6\la_7/\la_{45})b^2=-2\mu^2_1+\mu^2_3 \la_6/\la_{45},\label{ad4l}\\
&&(2\la_3-\la_6\la_7/\la_{45})a^2+(4\la_2-\la^2_7/\la_{45})b^2=-2\mu^2_2+\mu^2_3 \la_7/\la_{45}.\label{ad5l} \eea  
These equations always give a general solution for $a,b$ as supplied in terms of $v,w,v',w'$ from the outset in (\ref{29dd}). Additionally, since the $B-L$ violating potential parameters are small, $\bar{\la}\ll \la$ and $\bar{\mu}\ll \mu$, the equation (\ref{ad3l}) reveals $t=(vv'+w w')/\sqrt{(v^2+w'^2)(v'^2+w^2)}\ll 1$, in agreement with the constraints $v',w'\ll w,v$ in the body text.

Now we show that $\langle \rho\rangle,\langle \chi\rangle$ given above is a stable vacuum. Calculate 
\bea && \fr{\pa^2 V}{\pa t^2} =  2\la_{4}a^2 b^2,\\
&& \fr{\pa^2 V}{\pa t\pa a }= 2\la_{4}a b^2t + 2 \la_6 a^2 b,\\
&& \fr{\pa^2 V}{\pa t\pa b }= 2\la_{4}a^2 b t + 2 \la_7 a b^2,\\
&&\fr{\pa^2 V}{\pa a \pa b }= 4 \la_3 a b + 2\la_{4}a b t^2 + 2 \la_6 a^2 t + 2 \la_7 b^2 t,\\
&&\fr{\pa^2 V}{\pa a^2}= 8 \la_1 a^2 + 2\la_{4} b^2 t^2 + 4 \la_6 ab t,\\
&&\fr{\pa^2 V}{\pa b^2}= 8 \la_2 b^2 + 2\la_{4} a^2 t^2 + 4 \la_7 ab t.
 \eea The deviation of the potential from the extremum value is given by the Taylor's expansion up to second orders,
 \bea \Delta V &=& \fr{\pa^2 V}{\pa t^2} dt^2+\fr{\pa^2 V}{\pa a^2} da^2+\fr{\pa^2 V}{\pa b^2} db^2+2  \fr{\pa^2 V}{\pa t\pa a }dt da + 2  \fr{\pa^2 V}{\pa t\pa b } dt db +  2 \fr{\pa^2 V}{\pa a \pa b } d a d b\crn
&\geq& 8 \la_1 a^2 da^2 + 8 \la_2 b^2 db^2 + 8 \la_3 a b da db + 2 \la_4 a^2 b^2 dt^2,\label{ad6l} \eea where note that the first-order terms vanish due to the extremum conditions and that the last inequality (\ref{ad6l}) results due to the conditions, $t\ll 1, \bar{\la}\ll \la$, and $\la_4>0$ due to $m^2_{\mathcal{H,A}}>0$. The conditions for the potential bounded from below, i.e. $\la_{1,2}>0$, and positive Higgs squared-masses, i.e. $|\la_3|<2\sqrt{\la_1\la_2}$ and $\la_4>0$, lead to 
\bea \Delta V \geq 8\left( \sqrt{\la_1} a da + \fr{\la_3}{2\sqrt{\la_1}} b db \right)^2 +\fr{2}{\la_1}\left(4\la_1\la_2-\la^2_3\right) b^2 db^2 + 2 \la_4 a^2 b^2 dt^2>0, \eea which yields a minimum at $\langle \rho,\chi\rangle$. This minimum is global, i.e. vacuum stability, since the potential once bounded from below tends to $V\rightarrow +\infty$ for $a,b\rightarrow +\infty$. 

Last, but not least, expanding the scalar fields around the VEVs $\langle \rho,\chi\rangle$ as small fluctuations, the scalar potential takes the form $V(\rho,\chi) \simeq V({\langle\rho\rangle,\langle \chi\rangle}) + V_{\mathrm{mass}}$, where the scalar self interactions negligibly contribute as perturbative, and $V({\langle\rho\rangle,\langle \chi\rangle})$ is the potential value as in (\ref{addp}). Since $V_{\mathrm{mass}}$ is definitely positive, it follows that $V(\rho,\chi)>V({\langle\rho\rangle,\langle \chi\rangle})$, confirming again the stable minimum, which is also the true vacuum since $V\rightarrow +\infty$ for $\chi,\rho\rightarrow \infty$.      

\section{\label{appb} Gauge rotation}

Usually, an $SU(3)$ potential with two triplets $\phi_1$ and $\phi_2$, the VEVs can be brought to the form $\langle \phi_1\rangle=\fr{1}{\sqrt{2}} (0,0,v_1)^T$ and $\langle \phi_2 \rangle = \fr{1}{\sqrt{2}}(0,v_2,v_3)^T$ by a gauge rotation. And, the question that arises is how this is avoided here? The answer is simple: all the results and conclusions remain unchanged, since the theory is invariant under such gauge transformation. The difference in physics is only if which vacuum alignment, either [$\langle \chi\rangle = \fr{1}{\sqrt{2}}(0,v',w)^T$ and $\langle \rho\rangle = \fr{1}{\sqrt{2}}(0,v,w')^T$] or [$\langle \phi_1\rangle=\fr{1}{\sqrt{2}}(0,0,v_1)^T$ and $\langle \phi_2\rangle=\fr{1}{\sqrt{2}}(0,v_2,v_3)^T$], is chosen/considered from the beginning, where the basis of representation space has already been fixed by the quark and lepton arrangements.   

For details of the above judgement, the relevant gauge transformation is \be (0,v',w)^T = e^{i(\kappa /v_1)\la_7}(0,0,v_1)^T,\ee where 
\be e^{i(\kappa /v_1)\la_7}=\left(
\begin{array}{ccc}
1 & 0 & 0\\
0 & \cos\fr{\kappa}{v_1}& \sin\fr{\kappa}{v_1}\\
0 & -\sin\fr{\kappa}{v_1} & \cos\fr{\kappa}{v_1}
\end{array}\right),\ee $v' \equiv v_1\sin \fr{\kappa}{v_1}=\kappa -\fr 1 6 \fr{\kappa^3}{v^2_1}+\cdots\simeq \kappa$, and $w\equiv v_1\cos \fr{\kappa}{v_1}=v_1-\fr 1 2 \fr{\kappa^2}{v_1}+\cdots \simeq v_1$, where we use the condition $\kappa\ll v_1$ which is equivalent to $v'\ll w$, as expected. Correspondingly, the second triplet VEV is transformed as \be (0, v_2,v_3)^T=e^{-i(\kappa /v_1)\la_7}(0, v, w')^T\simeq (0,v,w'+v'v/w)^T.\ee
In the new basis by the gauge transformation, $U=e^{-i(\kappa /v_1)\la_7}$, all the fields must be shifted correspondingly. For instance, the quark and lepton multiplets are transformed to 
\be 
Q'_{aL}=U^* Q_{aL}\simeq \left(\begin{array}{c}
d_{a} \\
-u_{a}-\fr{v'}{w} U_a\\
U_{a}-\fr{v'}{w}u_a
\end{array}\right)_L,\hs \psi'_{\al L}=U\psi_{\al L}\simeq \left(\begin{array}{c}
\nu_{\al} \\
e_{\al}-\fr{v'}{w} E_\al\\
E_{\al}+\fr{v'}{w}e_\al
\end{array}\right)_L, \ee and $\psi'_{1L}= U \psi_{1L} U^T$, whereas the right-handed fermion singlets remain unchanged. 

Since the Lagrangian is invariant under $U$, it takes the same form after transformation with multiplets primed. Consider, for instance, the Yukawa interactions of quarks, 
\bea \mathcal{L}_{\mathrm{Yukawa}} &\supset& h_{ab}^{U}\bar{Q}'_{aL}\chi'^* U_{bR}+h^u_{ab}\bar{Q}'_{aL}\rho'^{*}u_{bR}\crn
&&+s^u_{ab}\bar{Q}'_{aL}\chi'^{*}u_{bR}+s_{ab}^{U}\bar{Q}'_{aL}\rho'^* U_{bR}\crn
&&+ \fr{h^d_{ab}}{\La}\bar{Q}'_{aL}\chi'\rho' d_{bR}+H.c., \eea where note that $\chi'\equiv \phi_1$ and $\rho'\equiv \phi_2$. Substituting the VEVs, we obtain 
\bea \mathcal{L}_{\mathrm{Yukawa}} &\supset&\fr{1}{\sqrt{2}}\left[\left(h^U_{ab}w +s^U_{ab}w' \right)\bar{U}_{aL}-\left(h^U_{ab}v' +s^U_{ab}v\right)u_{aL}\right] U_{bR} \crn
&&+\fr{1}{\sqrt{2}}\left[(s^u_{ab}w+h^u_{ab}w')\bar{U}_{aL}-(s^u_{ab}v'+h^u_{ab}v) \bar{u}_{aL}\right]u_{bR}
\crn
&& - \fr{h^d_{ab}}{2\La}vw\bar{d}_{aL}d_{bR}+H.c., \eea which yields quark masses coinciding with those obtained in the body text (\ref{32dd}) and (\ref{33dd}). Note that we approximate only up to the first orders in $(v',w')/(v,w)$, so the second order contribution, say $v'w'/vw$, disappears in the above down-quark masses. Similarly, we can obtain those masses and mixings for leptons, scalars, and gauge bosons.  

Even if we work in the new basis, the conditions (\ref{515d}) for the $PQ$ operator still retain. Indeed, the $PQ$ operator now becomes $PQ'=UPQ U^\dagger$, which annihilates the $\chi'$ and $\rho'$ vacua. We deduce, \bea &&-\fr{a}{2}+\fr{\sqrt{3}b}{2}=0,\label{adpq0}\\ 
&&\left(-\fr{a}{2}+\fr{b}{2\sqrt{3}}+\fr{c}{3}+dH_\chi\right)\sin^2\fr{v'}{w}+\left(-\fr{b}{\sqrt{3}}+\fr{c}{3}+dH_\chi\right)\cos\fr{v'}{w}\left(2-\cos\fr{v'}{w}\right)=0,\label{adpq1}\\
&&\left(-\fr{a}{2}+\fr{b}{2\sqrt{3}}+\fr{c}{3}+dH_\chi\right)\cos\fr{v'}{w}\left(2-\cos\fr{v'}{w}\right)+\left(-\fr{b}{\sqrt{3}}+\fr{c}{3}+dH_\chi\right)\sin^2\fr{v'}{w}=0\label{adpq2}.\eea  The equations (\ref{adpq1}) and (\ref{adpq2}) lead to (\ref{515d}), while (\ref{adpq0}) is a difference of these conditions.         

To conclude, all the results and conclusions that are obtained are independent of gauge rotation of a definite vacuum alignment. The choice of different vacuum alignments in a given basis would lead to different consequences, respectively. The mixing effects between ordinary and exotic quarks, leptons, and bosons in the current model due to $\chi,\rho$ make it distinct from the ordinary version with $v',w'=0$ where such physical effects disappear. 

\section{\label{appc} Stability of hierarchies}         

Let us discuss the hierarchies: (i) $v'\ll w$ within the vacuum $\langle \chi\rangle = \fr{1}{\sqrt{2}}(0,v',w)^T$, even stronger $v'\ll v\ll w$, and (ii) $w'\ll v$ within the vacuum $\langle \rho \rangle= \fr{1}{\sqrt{2}}(0,v,w')^T$, against quantum corrections. Note that these hierarchies have been determined from the physical points of view for the model consistency. (For instance, the $u$-$U$ mixing may otherwise lead to a violation of CKM unitarity and dangerous tree-level FCNC \cite{Dong:2014wsa}.)  

Concerning the first hierarchy, note that $\chi^0_2=\fr{1}{\sqrt{2}}v'+G^0_Y$ carries a $B-L$ charge by one unit and is $W_P$-odd. If $B-L$ is exact symmetry, but not spontaneously broken, its VEV, i.e. $v'$, vanishes. Even if $B-L$ is spontaneously broken, but preserving the matter parity $W_P$, then $v'=0$. Hence, when $v'\neq 0$, it breaks both $B-L$ and $W_P$. We expect that the approximate $B-L$ (and $W_P$) symmetry prevents $v'$ naturally small, as constrained by the $B-L$ violating parameters in the body text. 

The stability mechanism for such tiny $v'$ is that we do not have any physical (massive) Higgs field associate to the $v'$ scale, since $G^0_Y$ is already a Goldstone boson eaten by the non-Hermitian $Y^0$ gauge boson, as seen from the equation (\ref{615d}). The quantum corrections to the mass of $G_Y$, thus the shift $\Delta v'$, should vanish, up to any order of the perturbative theory, which is the consequence of the gauge symmetry or the Goldstone theorem. The stabilization of $v'$ is therefore followed, a direct result of the Goldstone theorem.

Indeed, apart from the vanishing mass term, the quantum contributions to the quartic coupling can be evaluated as \cite{Coleman:1973jx} \be V\supset \la_{\mathrm{eff}}(\chi^*_2 \chi_2)^2+\mathrm{violating\ couplings},\label{ddt1}\ee where
\be \la_{\mathrm{eff}}= \la_2 +\fr{A}{64\pi^2} \ln\fr{\chi^*_2 \chi_2}{\La^2} +\cdots\ee summarizes the tree value $\la_2$ and the one-loop corrections given in terms of $A=16\la^2_2+\la^2_3-2(h^E)^4-2(h^U)^4+3(1+1/4c^4_W)g^4$, which comes from the mediation of scalars, fermions, and gauge bosons that couple to $\chi_2$, respectively. Here $\La$ is an arbitrary renormalization scale at which $d^4V/d\chi^2_2 d\chi^{*2}_2=4\la_2$. Further, the dots represent higher order corrections which we expect they are in higher powers of $\ln \chi_2^2/\La^2$. Note that the violating potential slightly contribute to the quartic coupling, as neglected. All the log-terms are very insensitive to $\chi^2_2/\La^2$, approximately preserving a scale-symmetry at the regime of interest. These contributions should be radically smaller than the tree value $\la_2$ and we require $\la_{\mathrm{eff}}>0$ in order for the resulting potential bounded from below. The dominant part of (\ref{ddt1}) yields a minimum at $v'=0$, while the small value of $v'$ is lifted by the violating potential.    

Concerning the second hierarchy, note that $\rho^0_3=\fr{1}{\sqrt{2}}w'+H'$ carries a $B-L$ charge by minus-one unit and is $W_P$-odd, analogous to $\chi^0_2$. The $B-L$ and $W_P$ symmetries yield $w'=0$, while the approximate $B-L$ and $W_P$ symmetry, i.e. the violating potential, prevents $w'$ to be appropriately small. But, how does it work even at quantum levels? 

Observe that the tree-level potential of $\rho_3$ takes the form,
\be V\supset m^2_{H'}\rho^*_3 \rho_3+\la_1(\rho^*_3 \rho_3)^2+\mathrm{violating\ couplings},\ee where note that $\la_1>0$ and $m^2_{H'} \simeq \fr{\la_4}{2}(w^2+v^2)>0$ unlike the standard model scalar potential with a negative squared mass. The quantum corrections to $m^2_{H'}$ and $\la_1$ should be positive too. The reason for the positive effective coupling of $\rho_3$ as radiatively induced $\la'_{\mathrm{eff}}=\la_1+\Delta\la_1>0$ is analogous to that of $\chi_2$ ($\la_{\mathrm{eff}}$). Whereas, the contributions to the $H'$ mass depend on the cutoff scale, \be \Delta m^2_{H'}=\fr{4\la_1+3\la_3+\la_4}{16\pi^2}\La^2_{\mathrm{UV}},\ee which comes from the couplings of $\rho_{1,2}$ and $\chi_{1,2,3}$ to $\rho_3$, where the contribution of fermions and gauge bosons is small, as omitted. Since this theory is renormalizable, it demands that the radiatively-induced effective mass $m^2_{\mathrm{eff}}=m^2_{H'}+\Delta m^2_{H'}$ fixed at a renormalization scale should be finitely positive. 

Hence, we obtain 
\be V\supset m^2_{\mathrm{eff}}\rho^*_3 \rho_3+\la'_{\mathrm{eff}}(\rho^*_3 \rho_3)^2+\mathrm{violating\ couplings},\ee
where $m^2_{\mathrm{eff}}>0$ and $\la'_{\mathrm{eff}}>0$. The dominant potential, i.e. the first two terms, does not lead to spontaneous symmetry breaking, giving a minimum at $w'=0$. The small value of $w'$ is lifted, proportional to the violating parameters, but divided by the effective mass and coupling, as explicitly shown in the body text. The stability mechanism for $w'$ is just behind the scalar theory with positive squared-mass parameter and large quartic coupling, which ensures a stable minimum at $w'=0$, in agreement to \cite{Coleman:1973jx}. The large radiative modifications to $w'$ directly translating to the scalar effective mass and coupling term as inversely proportional to $w'$, making this $w'$ suitably tiny when including the violating potential.         

Last, the violating potential is important to set the size of $w',v'$ as partly discussed below.      

\section{\label{appd} Strength of approximate symmetry violation}

\subsection{\label{appd1} Implication from top-down approach}
As shown in the body text, for the 3-3-1 model, if $B-L$ is conserved it must be a gauged charge, requiring a 3-3-1-1 extension, since $B-L$ neither commutes nor closes algebraically with $SU(3)_L$. In such case, all the interactions and mass terms that explicitly violate $B-L$ must be absent. In the considering model, the $B-L$ violating couplings existing in the Yukawa and scalar sectors, say $s$, $s'$, $\bar{\mu}$, and $\bar{\la}$, measure an approximate $B-L$ symmetry, required in order for the 3-3-1 model to be self-consistent. We expect that a number of such $B-L$ violating couplings are imprinted from the 3-3-1-1 symmetry breaking down to 3-3-1, after integrating out the heavy particles of extended symmetry such as the $U(1)_N$ gauge boson, right-handed neutrinos and Higgs scalars \cite{Dong:2013wca,Dong:2014wsa,Huong:2015dwa,Dong:2015yra,Huong:2016ybt,Alves:2016fqe,Dong:2018aak}. This would provide an appropriate estimation of the violating strength of the approximate symmetry.

In the current model, the strength of the $B-L$ violating effective interactions responsible for neutrino masses is typically proportional to $s'/\La$, such as \be \fr{s'}{\La}\psi_{1L}\psi_{1L}\rho\rho,\label{ssddtt}\ee where note that $\La\sim \langle\chi\rangle\sim 10$ TeV is given as in the body text. On the other hand, extended to the 3-3-1-1 model, both $\psi_{1L}$ and $\rho$ may couple to a heavy scalar sextet $\sigma\sim(1,6,2/3,4/3)$ through $h \psi_{1L}\psi_{1L}\sigma$ and $\la \phi \rho^T \sigma^*\rho$, where $\phi\sim (1,1,0,2)$ is necessary to break $U(1)_N$ and generate right-handed neutrino masses. This induces an effective interaction, \be \fr{h\la}{\La_N}\psi_{1L}\psi_{1L}\rho\rho,\label{extrdd}\ee after breaking $U(1)_N$ by $\La_N=\langle \phi\rangle$ and integrating $\sigma$ out with a mass at $\La_N$ scale. This vacuum structure of the heavy scalar sector with potential $V(\phi,\sigma)=\mu^2_\phi \phi^\dagger \phi+\mu^2_\sigma\Tr \sigma^\dagger \sigma+\la_\phi (\phi^\dagger \phi)^2 +\la_\sigma \Tr^2 \sigma^\dagger \sigma+\la'_\sigma \Tr(\sigma^\dagger \sigma)^2+\la_{\phi \sigma}\phi^\dagger \phi (\Tr \sigma^\dagger \sigma)$ can be achieved by choosing the parameters such that $\mu^2_\phi<0$, $\mu^2_\sigma>0$, $\la_{\phi}>0$, $\la_\sigma+\la'_\sigma>0$, and $\la_{\phi \sigma}>-2[\la_\phi (\la_\sigma+\la_\sigma')]^{1/2}$, thus $\La_N=(-\mu^2_\phi/2\la_\phi)^{1/2}$, $\langle \sigma\rangle =0$, and $m^2_\sigma=\mu^2_\sigma+\la_{\phi \sigma}\La^2_N\sim \La^2_N$.    

In a toy model mentioned in the body text, $\rho$ is replaced by $\eta$ that now couples to $h_{\al b}\bar{\psi}_{\al L}\eta \nu_{bR}$. Recall that $\nu_R$ gain heavy Majorana masses via the coupling $\fr 1 2 f_{ab}\bar{\nu}^c_{aR}\phi\nu_{bR}$ to be $m_{\nu_R}=-f\La_N$. Integrating $\nu_R$ out, we obtain an effective interaction, \be \fr{c_{\al\beta}}{\La_N}\psi_{\al L}\psi_{\beta L}\eta^*\eta^*,\label{ddtnl}\ee where $c\equiv hf^{-1}h^T$. Both (\ref{extrdd}) and (\ref{ddtnl}) are Weinberg-type effective operators since $\rho$ and $\eta$ contain standard model-like Higgs doublets. The disappearance of $\eta$ suppresses relevant effective interactions, but necessarily implies that the effect of the 3-3-1-1 breaking translates to the low energy theory the same strength $\sim 1/\La_N$, where it is suitably to take $c\sim 1\sim h\la$. Comparing to (\ref{ssddtt}), it matches $\fr{s'}{\La}\sim \fr{1}{\La_N}$, which follows that 
\be s'\sim \fr{\La}{\La_N}\sim 10^{-10},\ee where we take $\La_N\sim 10^{14}$ GeV coinciding with the inflation scale derived by the $U(1)_N$ dynamics \cite{Huong:2015dwa,Huong:2016ybt,Dong:2018aak}. Such $s'$ value agrees with the neutrino mass constraint in the body text. 

Hence, the neutrino masses are partly implied by $B-L=2$ violating effective interactions originating from the 3-3-1-1 breakdown by the $\phi$ field. However, the $\phi$ vacuum conserves the matter parity, $W_P \La_N=\La_N$, which imprints at the low energy the 3-3-1 and $W_P$ symmetries. All the other interactions in Yukawa Lagrangian and scalar potential that violate $B-L$ by one unit are suppressed by $W_P$, since such couplings, $\chi^\dagger \rho$, $\rho^\dagger\rho\chi^\dagger \rho$, $\bar{Q}_L \chi^* u_R$, $\bar{\psi}_{\al L}\chi e_{bR}$, $\psi_{1L}\chi\rho\psi_{1L}$, $\psi_{1L}\chi\chi\rho\psi_{\al L}$, and so forth, are $W_P$ odd. They may be generated by a violation of $W_P$ in the effective theory, but should be comparable to the above neutrino mass operators, where particularly the $B-L=1$ violating interactions are required to produce completely consistent neutrino masses.   

\subsection{\label{appd2} Stability of violating couplings}

The stability of violating effective Yukawa couplings is obvious, as can be seen from the realm of renormalization theory as well as the above discussion. Indeed, these effective interactions do not have any tree-level origin at renormalizable level (in other words, their bare coupling constant or corresponding counterterm vanish). Hence, if these effective couplings are radiatively induced by the model particles, they should be finite, but strongly suppressed by a loop factor and mass scale, e.g. $1/16\pi^2\La$, multiplied by the contribution of the $B-L$ violating renormalizable Lagrangian (since the $B-L$ conservation suppresses them up to any order of perturbation theory). With the aid of dimensional analysis and $B-L$ number count, we evaluate \be s\sim \fr{1}{16\pi^2}\fr{v',w'}{\La},\hs s'\sim \fr{1}{16\pi^2}\left(\fr{v',w'}{\La}\right)^2,\ee for breaking $B-L$ by one and two units, respectively. Taking the maximal value $v',w'\sim 0.1$ MeV as given below (\ref{67d}), we derive $s\sim 10^{-10}$ and $s'\sim 10^{-18}$. Appropriate to the neutrino mass constraint, $s\sim 10^{-6}$ and $s'\sim 10^{-10}$, one must take $v',w'\sim 1$ GeV which are unlikely in the current model. Thus, the radiative corrections by the model particles do not destroy the violating effective couplings implied by approximate $B-L$ symmetry. On the other hand, the running of such couplings up to a high energy scale, the new physics enters, inducing them as suppressed by a seesaw scale---the $B-L$ breaking energy---with suitable values as determined above from the top-down approach.   

\begin{figure}[h]
\bc
\includegraphics[scale=0.7]{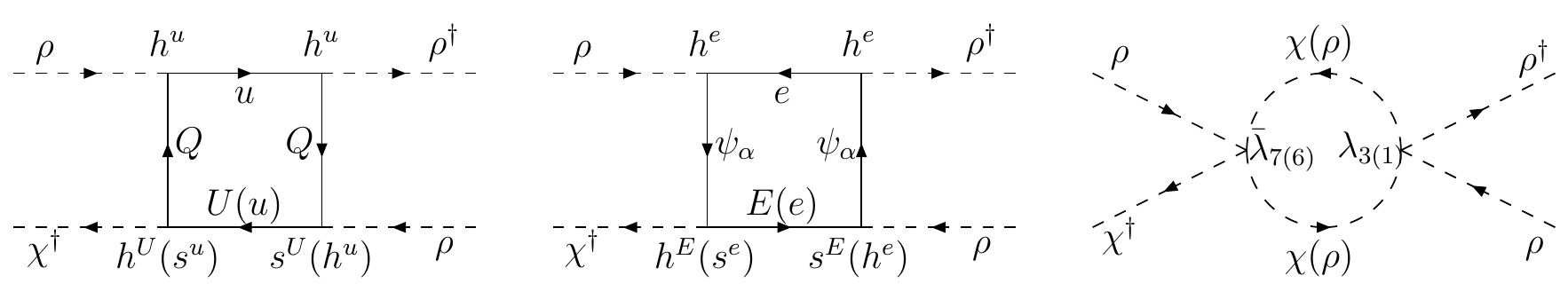}
\caption[]{\label{addfig}Dominant one-loop contributions to the $\bar{\la}_6$ coupling, where each diagram-type has an additional contribution with parenthesized fields and couplings.} 
\ec
\end{figure}

To investigate the stability of $B-L$ violating renormalizable Yukawa and scalar self couplings, we consider only $\bar{\la}_6$. The other couplings of such kinds can be achieved similarly. The one-loop corrections to $\bar{\la}_6$ are given in Fig. \ref{addfig}.
Recall that although the neutrino masses are generated, the low energy theory obeys a good residual symmetry, the matter parity $W_P$, where the violating renormalizable couplings are $W_P$ odd. Such couplings vanish up to any order of perturbation theory, if $W_P$ is preserved. Hence, the one-loop corrections to $\bar{\la}_6$ are necessarily odd under $W_P$ as $\bar{\la}_{6}$ is. This can be verified directly from the above diagrams, where $s^{U,u}$, $s^{E,e}$ and $\bar{\la}_7$ that enter are all $W_P$ odd. Since the ultraviolet divergences can be absorbed into the bare coupling by renormalization condition $d^4 V/d\chi^\dagger d \rho d \rho^\dagger d \rho = 2\bar{\la}_6$, the one-loop coupling strength takes the form,
\bea \bar{\la}^{\mathrm{eff}}_6 &=&\bar{\la}_6+ \fr{1}{16\pi^2}\left(h^U s^{U\dagger} h^u h^{u\dagger} F_1+s^u h^{u\dagger} h^u h^{u\dagger} F_2 \right.\crn
&&\left.+h^{E\dagger} s^{E}  h^{e\dagger}h^e F_3+s^{e\dagger} h^{e}  h^{e\dagger}h^e F_4+\bar{\la}_7\la_3 F_5+\bar{\la}_6\la_1F_6\right),\eea where $F_{1,2}$, $F_{3,4}$ and $F_{5,6}$ are finite functions of quark, lepton, and scalar masses divided by renormalization mass, respectively, which need not necessarily be determined. It is clear that the radiative corrections to $\bar{\la}_6$ are strongly suppressed by the corresponding violating couplings, if one takes $s^{U}\sim s^{u}\sim s^E\sim s^e\sim \bar{\la}_7\sim \bar{\la}_6$. From this view, we cannot retain some violating coupling, e.g. $\bar{\la}_{6}$, to be much smaller (i.e. large hierarchical) than the other couplings of this type. However, all of them must be as small as the approximate matter parity is allowed. Again using the top-down approach, if $B-L$ is completely broken, we expect that $\bar{\la}_6$ is similar to that for neutrino mass generation.

A detailed study on all hierarchies including the couplings and VEVs is interesting, but it is out of the scope of this work, to be published elsewhere.     

\bibliographystyle{JHEP}
\bibliography{combine}

\providecommand{\href}[2]{#2}\begingroup\raggedright\begin{thebibliography}{10}

\bibitem{Pisano:1991ee}
F.~Pisano and V.~Pleitez, \emph{{An SU(3) x U(1) model for electroweak
  interactions}}, \href{https://doi.org/10.1103/PhysRevD.46.410}{\emph{Phys.
  Rev.} {\bfseries D46} (1992) 410}
  [\href{https://arxiv.org/abs/hep-ph/9206242}{{\ttfamily hep-ph/9206242}}].

\bibitem{Frampton:1992wt}
P.~H. Frampton, \emph{{Chiral dilepton model and the flavor question}},
  \href{https://doi.org/10.1103/PhysRevLett.69.2889}{\emph{Phys. Rev. Lett.}
  {\bfseries 69} (1992) 2889}.

\bibitem{Foot:1992rh}
R.~Foot, O.~F. Hernandez, F.~Pisano and V.~Pleitez, \emph{{Lepton masses in an
  SU(3)-L x U(1)-N gauge model}},
  \href{https://doi.org/10.1103/PhysRevD.47.4158}{\emph{Phys. Rev.} {\bfseries
  D47} (1993) 4158} [\href{https://arxiv.org/abs/hep-ph/9207264}{{\ttfamily
  hep-ph/9207264}}].

\bibitem{Singer:1980sw}
M.~Singer, J.~W.~F. Valle and J.~Schechter, \emph{{Canonical Neutral Current
  Predictions From the Weak Electromagnetic Gauge Group SU(3) X $u$(1)}},
  \href{https://doi.org/10.1103/PhysRevD.22.738}{\emph{Phys. Rev.} {\bfseries
  D22} (1980) 738}.

\bibitem{Montero:1992jk}
J.~C. Montero, F.~Pisano and V.~Pleitez, \emph{{Neutral currents and GIM
  mechanism in SU(3)-L x U(1)-N models for electroweak interactions}},
  \href{https://doi.org/10.1103/PhysRevD.47.2918}{\emph{Phys. Rev.} {\bfseries
  D47} (1993) 2918} [\href{https://arxiv.org/abs/hep-ph/9212271}{{\ttfamily
  hep-ph/9212271}}].

\bibitem{Foot:1994ym}
R.~Foot, H.~N. Long and T.~A. Tran, \emph{{$SU(3)_L \otimes U(1)_N$ and
  $SU(4)_L \otimes U(1)_N$ gauge models with right-handed neutrinos}},
  \href{https://doi.org/10.1103/PhysRevD.50.R34}{\emph{Phys. Rev.} {\bfseries
  D50} (1994) R34} [\href{https://arxiv.org/abs/hep-ph/9402243}{{\ttfamily
  hep-ph/9402243}}].

\bibitem{Pisano:1996ht}
F.~Pisano, \emph{{A Simple solution for the flavor question}},
  \href{https://doi.org/10.1142/S0217732396002630}{\emph{Mod. Phys. Lett.}
  {\bfseries A11} (1996) 2639}
  [\href{https://arxiv.org/abs/hep-ph/9609358}{{\ttfamily hep-ph/9609358}}].

\bibitem{Doff:1998we}
A.~Doff and F.~Pisano, \emph{{Charge quantization in the largest leptoquark
  bilepton chiral electroweak scheme}},
  \href{https://doi.org/10.1142/S0217732399001218}{\emph{Mod. Phys. Lett.}
  {\bfseries A14} (1999) 1133}
  [\href{https://arxiv.org/abs/hep-ph/9812303}{{\ttfamily hep-ph/9812303}}].

\bibitem{deSousaPires:1998jc}
C.~A. de~Sousa~Pires and O.~P. Ravinez, \emph{{Charge quantization in a chiral
  bilepton gauge model}},
  \href{https://doi.org/10.1103/PhysRevD.58.035008}{\emph{Phys. Rev.}
  {\bfseries D58} (1998) 035008}
  [\href{https://arxiv.org/abs/hep-ph/9803409}{{\ttfamily hep-ph/9803409}}].

\bibitem{deSousaPires:1999ca}
C.~A. de~Sousa~Pires, \emph{{Remark on the vector - like nature of the
  electromagnetism and the electric charge quantization}},
  \href{https://doi.org/10.1103/PhysRevD.60.075013}{\emph{Phys. Rev.}
  {\bfseries D60} (1999) 075013}
  [\href{https://arxiv.org/abs/hep-ph/9902406}{{\ttfamily hep-ph/9902406}}].

\bibitem{VanDong:2005ux}
P.~V. Dong and H.~N. Long, \emph{{Electric charge quantization in SU(3)(C) x
  SU(3)(L) x U(1)(X) models}},
  \href{https://doi.org/10.1142/S0217751X06035191}{\emph{Int. J. Mod. Phys.}
  {\bfseries A21} (2006) 6677}
  [\href{https://arxiv.org/abs/hep-ph/0507155}{{\ttfamily hep-ph/0507155}}].

\bibitem{Pal:1994ba}
P.~B. Pal, \emph{{The Strong CP question in SU(3)(C) x SU(3)(L) x U(1)(N)
  models}}, \href{https://doi.org/10.1103/PhysRevD.52.1659}{\emph{Phys. Rev.}
  {\bfseries D52} (1995) 1659}
  [\href{https://arxiv.org/abs/hep-ph/9411406}{{\ttfamily hep-ph/9411406}}].

\bibitem{Dong:2012bf}
P.~V. Dong, H.~N. Long and H.~T. Hung, \emph{{Question of Peccei-Quinn symmetry
  and quark masses in the economical 3-3-1 model}},
  \href{https://doi.org/10.1103/PhysRevD.86.033002}{\emph{Phys. Rev.}
  {\bfseries D86} (2012) 033002}
  [\href{https://arxiv.org/abs/1205.5648}{{\ttfamily 1205.5648}}].

\bibitem{Tully:2000kk}
M.~B. Tully and G.~C. Joshi, \emph{{Generating neutrino mass in the 331
  model}}, \href{https://doi.org/10.1103/PhysRevD.64.011301}{\emph{Phys. Rev.}
  {\bfseries D64} (2001) 011301}
  [\href{https://arxiv.org/abs/hep-ph/0011172}{{\ttfamily hep-ph/0011172}}].

\bibitem{Dias:2005yh}
A.~G. Dias, C.~A. de~S.~Pires and P.~S. Rodrigues~da Silva, \emph{{Naturally
  light right-handed neutrinos in a 3-3-1 model}},
  \href{https://doi.org/10.1016/j.physletb.2005.09.028}{\emph{Phys. Lett.}
  {\bfseries B628} (2005) 85}
  [\href{https://arxiv.org/abs/hep-ph/0508186}{{\ttfamily hep-ph/0508186}}].

\bibitem{Chang:2006aa}
D.~Chang and H.~N. Long, \emph{{Interesting radiative patterns of neutrino mass
  in an SU(3)(C) x SU(3)(L) x U(1)(X) model with right-handed neutrinos}},
  \href{https://doi.org/10.1103/PhysRevD.73.053006}{\emph{Phys. Rev.}
  {\bfseries D73} (2006) 053006}
  [\href{https://arxiv.org/abs/hep-ph/0603098}{{\ttfamily hep-ph/0603098}}].

\bibitem{Dong:2006mt}
P.~V. Dong, H.~N. Long and D.~V. Soa, \emph{{Neutrino masses in the economical
  3-3-1 model}}, \href{https://doi.org/10.1103/PhysRevD.75.073006}{\emph{Phys.
  Rev.} {\bfseries D75} (2007) 073006}
  [\href{https://arxiv.org/abs/hep-ph/0610381}{{\ttfamily hep-ph/0610381}}].

\bibitem{Dong:2008sw}
P.~V. Dong and H.~N. Long, \emph{{Neutrino masses and lepton flavor violation
  in the 3-3-1 model with right-handed neutrinos}},
  \href{https://doi.org/10.1103/PhysRevD.77.057302}{\emph{Phys. Rev.}
  {\bfseries D77} (2008) 057302}
  [\href{https://arxiv.org/abs/0801.4196}{{\ttfamily 0801.4196}}].

\bibitem{Dong:2010gk}
P.~V. Dong, L.~T. Hue, H.~N. Long and D.~V. Soa, \emph{{The 3-3-1 model with
  $A_4$ flavor symmetry}},
  \href{https://doi.org/10.1103/PhysRevD.81.053004}{\emph{Phys. Rev.}
  {\bfseries D81} (2010) 053004}
  [\href{https://arxiv.org/abs/1001.4625}{{\ttfamily 1001.4625}}].

\bibitem{Dong:2010zu}
P.~V. Dong, H.~N. Long, D.~V. Soa and V.~V. Vien, \emph{{The 3-3-1 model with
  $S_4$ flavor symmetry}},
  \href{https://doi.org/10.1140/epjc/s10052-011-1544-2}{\emph{Eur. Phys. J.}
  {\bfseries C71} (2011) 1544}
  [\href{https://arxiv.org/abs/1009.2328}{{\ttfamily 1009.2328}}].

\bibitem{Dong:2011vb}
P.~V. Dong, H.~N. Long, C.~H. Nam and V.~V. Vien, \emph{{The $S_3$ flavor
  symmetry in 3-3-1 models}},
  \href{https://doi.org/10.1103/PhysRevD.85.053001}{\emph{Phys. Rev.}
  {\bfseries D85} (2012) 053001}
  [\href{https://arxiv.org/abs/1111.6360}{{\ttfamily 1111.6360}}].

\bibitem{Dias:2012xp}
A.~G. Dias, C.~A. de~S.~Pires, P.~S. Rodrigues~da Silva and A.~Sampieri,
  \emph{{A Simple Realization of the Inverse Seesaw Mechanism}},
  \href{https://doi.org/10.1103/PhysRevD.86.035007}{\emph{Phys. Rev.}
  {\bfseries D86} (2012) 035007}
  [\href{https://arxiv.org/abs/1206.2590}{{\ttfamily 1206.2590}}].

\bibitem{Catano:2012kw}
M.~E. Catano, R.~Martinez and F.~Ochoa, \emph{{Neutrino masses in a 331 model
  with right-handed neutrinos without doubly charged Higgs bosons via inverse
  and double seesaw mechanisms}},
  \href{https://doi.org/10.1103/PhysRevD.86.073015}{\emph{Phys. Rev.}
  {\bfseries D86} (2012) 073015}
  [\href{https://arxiv.org/abs/1206.1966}{{\ttfamily 1206.1966}}].

\bibitem{Boucenna:2014ela}
S.~M. Boucenna, S.~Morisi and J.~W.~F. Valle, \emph{{Radiative neutrino mass in
  3-3-1 scheme}}, \href{https://doi.org/10.1103/PhysRevD.90.013005}{\emph{Phys.
  Rev.} {\bfseries D90} (2014) 013005}
  [\href{https://arxiv.org/abs/1405.2332}{{\ttfamily 1405.2332}}].

\bibitem{Boucenna:2014dia}
S.~M. Boucenna, R.~M. Fonseca, F.~Gonzalez-Canales and J.~W.~F. Valle,
  \emph{{Small neutrino masses and gauge coupling unification}},
  \href{https://doi.org/10.1103/PhysRevD.91.031702}{\emph{Phys. Rev.}
  {\bfseries D91} (2015) 031702}
  [\href{https://arxiv.org/abs/1411.0566}{{\ttfamily 1411.0566}}].

\bibitem{Boucenna:2015zwa}
S.~M. Boucenna, J.~W.~F. Valle and A.~Vicente, \emph{{Predicting charged lepton
  flavor violation from 3-3-1 gauge symmetry}},
  \href{https://doi.org/10.1103/PhysRevD.92.053001}{\emph{Phys. Rev.}
  {\bfseries D92} (2015) 053001}
  [\href{https://arxiv.org/abs/1502.07546}{{\ttfamily 1502.07546}}].

\bibitem{Okada:2015bxa}
H.~Okada, N.~Okada and Y.~Orikasa, \emph{{Radiative seesaw mechanism in a
  minimal 3-3-1 model}},
  \href{https://doi.org/10.1103/PhysRevD.93.073006}{\emph{Phys. Rev.}
  {\bfseries D93} (2016) 073006}
  [\href{https://arxiv.org/abs/1504.01204}{{\ttfamily 1504.01204}}].

\bibitem{Pires:2014xsa}
C.~A. d.~S. Pires, \emph{{Neutrino mass mechanisms in 3-3-1 models: A short
  review}},  \href{https://arxiv.org/abs/1412.1002}{{\ttfamily 1412.1002}}.

\bibitem{Fregolente:2002nx}
D.~Fregolente and M.~D. Tonasse, \emph{{Selfinteracting dark matter from an
  SU(3)(L) x U(1)(N) electroweak model}},
  \href{https://doi.org/10.1016/S0370-2693(03)00037-6}{\emph{Phys. Lett.}
  {\bfseries B555} (2003) 7}
  [\href{https://arxiv.org/abs/hep-ph/0209119}{{\ttfamily hep-ph/0209119}}].

\bibitem{Hoang:2003vj}
H.~N. Long and N.~Q. Lan, \emph{{Selfinteracting dark matter and Higgs bosons
  in the SU(3)(C) x SU(3)(L) x U(1)(N) model with right-handed neutrinos}},
  \href{https://doi.org/10.1209/epl/i2003-00267-5}{\emph{Europhys. Lett.}
  {\bfseries 64} (2003) 571}
  [\href{https://arxiv.org/abs/hep-ph/0309038}{{\ttfamily hep-ph/0309038}}].

\bibitem{Filippi:2005mt}
S.~Filippi, W.~A. Ponce and L.~A. Sanchez, \emph{{Dark matter from the scalar
  sector of 3-3-1 models without exotic electric charges}},
  \href{https://doi.org/10.1209/epl/i2005-10349-x}{\emph{Europhys. Lett.}
  {\bfseries 73} (2006) 142}
  [\href{https://arxiv.org/abs/hep-ph/0509173}{{\ttfamily hep-ph/0509173}}].

\bibitem{deS.Pires:2007gi}
C.~A. de~S.~Pires and P.~S. Rodrigues~da Silva, \emph{{Scalar Bilepton Dark
  Matter}}, \href{https://doi.org/10.1088/1475-7516/2007/12/012}{\emph{JCAP}
  {\bfseries 0712} (2007) 012}
  [\href{https://arxiv.org/abs/0710.2104}{{\ttfamily 0710.2104}}].

\bibitem{Mizukoshi:2010ky}
J.~K. Mizukoshi, C.~A. de~S.~Pires, F.~S. Queiroz and P.~S. Rodrigues~da Silva,
  \emph{{WIMPs in a 3-3-1 model with heavy Sterile neutrinos}},
  \href{https://doi.org/10.1103/PhysRevD.83.065024}{\emph{Phys. Rev.}
  {\bfseries D83} (2011) 065024}
  [\href{https://arxiv.org/abs/1010.4097}{{\ttfamily 1010.4097}}].

\bibitem{Alvares:2012qv}
J.~D. Ruiz-Alvarez, C.~A. de~S.~Pires, F.~S. Queiroz, D.~Restrepo and P.~S.
  Rodrigues~da Silva, \emph{{On the Connection of Gamma-Rays, Dark Matter and
  Higgs Searches at LHC}},
  \href{https://doi.org/10.1103/PhysRevD.86.075011}{\emph{Phys. Rev.}
  {\bfseries D86} (2012) 075011}
  [\href{https://arxiv.org/abs/1206.5779}{{\ttfamily 1206.5779}}].

\bibitem{Profumo:2013sca}
S.~Profumo and F.~S. Queiroz, \emph{{Constraining the $Z'$ mass in 331 models
  using direct dark matter detection}},
  \href{https://doi.org/10.1140/epjc/s10052-014-2960-x}{\emph{Eur. Phys. J.}
  {\bfseries C74} (2014) 2960}
  [\href{https://arxiv.org/abs/1307.7802}{{\ttfamily 1307.7802}}].

\bibitem{Kelso:2013nwa}
C.~Kelso, C.~A. de~S.~Pires, S.~Profumo, F.~S. Queiroz and P.~S. Rodrigues~da
  Silva, \emph{{A 331 WIMPy Dark Radiation Model}},
  \href{https://doi.org/10.1140/epjc/s10052-014-2797-3}{\emph{Eur. Phys. J.}
  {\bfseries C74} (2014) 2797}
  [\href{https://arxiv.org/abs/1308.6630}{{\ttfamily 1308.6630}}].

\bibitem{daSilva:2014qba}
P.~S. Rodrigues~da Silva, \emph{{A Brief Review on WIMPs in 331 Electroweak
  Gauge Models}}, \href{https://doi.org/10.3844/pisp.2016.15.27}{\emph{Phys.
  Int.} {\bfseries 7} (2016) 15}
  [\href{https://arxiv.org/abs/1412.8633}{{\ttfamily 1412.8633}}].

\bibitem{Dong:2013ioa}
P.~V. Dong, T.~P. Nguyen and D.~V. Soa, \emph{{3-3-1 model with inert scalar
  triplet}}, \href{https://doi.org/10.1103/PhysRevD.88.095014}{\emph{Phys.
  Rev.} {\bfseries D88} (2013) 095014}
  [\href{https://arxiv.org/abs/1308.4097}{{\ttfamily 1308.4097}}].

\bibitem{Dong:2014esa}
P.~V. Dong, N.~T.~K. Ngan and D.~V. Soa, \emph{{Simple 3-3-1 model and
  implication for dark matter}},
  \href{https://doi.org/10.1103/PhysRevD.90.075019}{\emph{Phys. Rev.}
  {\bfseries D90} (2014) 075019}
  [\href{https://arxiv.org/abs/1407.3839}{{\ttfamily 1407.3839}}].

\bibitem{Dong:2015rka}
P.~V. Dong, C.~S. Kim, D.~V. Soa and N.~T. Thuy, \emph{{Investigation of Dark
  Matter in Minimal 3-3-1 Models}},
  \href{https://doi.org/10.1103/PhysRevD.91.115019}{\emph{Phys. Rev.}
  {\bfseries D91} (2015) 115019}
  [\href{https://arxiv.org/abs/1501.04385}{{\ttfamily 1501.04385}}].

\bibitem{Dong:2013wca}
P.~V. Dong, H.~T. Hung and T.~D. Tham, \emph{{3-3-1-1 model for dark matter}},
  \href{https://doi.org/10.1103/PhysRevD.87.115003}{\emph{Phys. Rev.}
  {\bfseries D87} (2013) 115003}
  [\href{https://arxiv.org/abs/1305.0369}{{\ttfamily 1305.0369}}].

\bibitem{Dong:2014wsa}
P.~V. Dong, D.~T. Huong, F.~S. Queiroz and N.~T. Thuy, \emph{{Phenomenology of
  the 3-3-1-1 model}},
  \href{https://doi.org/10.1103/PhysRevD.90.075021}{\emph{Phys. Rev.}
  {\bfseries D90} (2014) 075021}
  [\href{https://arxiv.org/abs/1405.2591}{{\ttfamily 1405.2591}}].

\bibitem{Dong:2015yra}
P.~V. Dong, \emph{{Unifying the electroweak and B-L interactions}},
  \href{https://doi.org/10.1103/PhysRevD.92.055026}{\emph{Phys. Rev.}
  {\bfseries D92} (2015) 055026}
  [\href{https://arxiv.org/abs/1505.06469}{{\ttfamily 1505.06469}}].

\bibitem{Huong:2016ybt}
D.~T. Huong and P.~V. Dong, \emph{{Neutrino masses and superheavy dark matter
  in the 3-3-1-1 model}},
  \href{https://doi.org/10.1140/epjc/s10052-017-4763-3}{\emph{Eur. Phys. J.}
  {\bfseries C77} (2017) 204}
  [\href{https://arxiv.org/abs/1605.01216}{{\ttfamily 1605.01216}}].

\bibitem{Alves:2016fqe}
A.~Alves, G.~Arcadi, P.~V. Dong, L.~Duarte, F.~S. Queiroz and J.~W.~F. Valle,
  \emph{{Matter-parity as a residual gauge symmetry: Probing a theory of
  cosmological dark matter}},
  \href{https://doi.org/10.1016/j.physletb.2017.07.056}{\emph{Phys. Lett.}
  {\bfseries B772} (2017) 825}
  [\href{https://arxiv.org/abs/1612.04383}{{\ttfamily 1612.04383}}].

\bibitem{Huong:2015dwa}
D.~T. Huong, P.~V. Dong, C.~S. Kim and N.~T. Thuy, \emph{{Inflation and
  leptogenesis in the 3-3-1-1 model}},
  \href{https://doi.org/10.1103/PhysRevD.91.055023}{\emph{Phys. Rev.}
  {\bfseries D91} (2015) 055023}
  [\href{https://arxiv.org/abs/1501.00543}{{\ttfamily 1501.00543}}].

\bibitem{Dong:2018aak}
P.~V. Dong, D.~T. Huong, D.~A. Camargo, F.~S. Queiroz and J.~W.~F. Valle,
  \emph{{Asymmetric Dark Matter, Inflation and Leptogenesis from B-L Symmetry
  Breaking}},  \href{https://arxiv.org/abs/1805.08251}{{\ttfamily 1805.08251}}.

\bibitem{Dong:2017ayu}
P.~V. Dong, D.~Q. Phong, D.~V. Soa and N.~C. Thao, \emph{{The economical 3-3-1
  model revisited}},  \href{https://arxiv.org/abs/1706.06152}{{\ttfamily
  1706.06152}}.

\bibitem{Ferreira:2016uao}
J.~G. Ferreira, C.~A. de~S.~Pires, J.~G. Rodrigues and P.~S. Rodrigues~da
  Silva, \emph{{Embedding cosmological inflation, axion dark matter and seesaw
  mechanism in a 3-3-1 gauge model}},
  \href{https://doi.org/10.1016/j.physletb.2017.05.034}{\emph{Phys. Lett.}
  {\bfseries B771} (2017) 199}
  [\href{https://arxiv.org/abs/1612.01463}{{\ttfamily 1612.01463}}].

\bibitem{Fonseca:2016tbn}
R.~M. Fonseca and M.~Hirsch, \emph{{A flipped 331 model}},
  \href{https://doi.org/10.1007/JHEP08(2016)003}{\emph{JHEP} {\bfseries 08}
  (2016) 003} [\href{https://arxiv.org/abs/1606.01109}{{\ttfamily
  1606.01109}}].

\bibitem{Gross:1972pv}
D.~J. Gross and R.~Jackiw, \emph{{Effect of anomalies on quasirenormalizable
  theories}}, \href{https://doi.org/10.1103/PhysRevD.6.477}{\emph{Phys. Rev.}
  {\bfseries D6} (1972) 477}.

\bibitem{Georgi:1972bb}
H.~Georgi and S.~L. Glashow, \emph{{Gauge theories without anomalies}},
  \href{https://doi.org/10.1103/PhysRevD.6.429}{\emph{Phys. Rev.} {\bfseries
  D6} (1972) 429}.

\bibitem{Banks:1976yg}
J.~Banks and H.~Georgi, \emph{{Comment on Gauge Theories Without Anomalies}},
  \href{https://doi.org/10.1103/PhysRevD.14.1159}{\emph{Phys. Rev.} {\bfseries
  D14} (1976) 1159}.

\bibitem{Okubo:1977sc}
S.~Okubo, \emph{{Gauge Groups Without Triangular Anomaly}},
  \href{https://doi.org/10.1103/PhysRevD.16.3528}{\emph{Phys. Rev.} {\bfseries
  D16} (1977) 3528}.

\bibitem{Ng:1992st}
D.~Ng, \emph{{The Electroweak theory of SU(3) x U(1)}},
  \href{https://doi.org/10.1103/PhysRevD.49.4805}{\emph{Phys. Rev.} {\bfseries
  D49} (1994) 4805} [\href{https://arxiv.org/abs/hep-ph/9212284}{{\ttfamily
  hep-ph/9212284}}].

\bibitem{GomezDumm:1994tz}
F.~Pisano, D.~Gomez~Dumm, F.~Pisano and V.~Pleitez, \emph{{Flavor changing
  neutral currents in SU(3) x U(1) models}},
  \href{https://doi.org/10.1142/S0217732394001441}{\emph{Mod. Phys. Lett.}
  {\bfseries A9} (1994) 1609}
  [\href{https://arxiv.org/abs/hep-ph/9307265}{{\ttfamily hep-ph/9307265}}].

\bibitem{Long:1999ij}
H.~N. Long and V.~T. Van, \emph{{Quark family discrimination and flavor
  changing neutral currents in the SU(3)(C) x SU(3)(L) x U(1) model with
  right-handed neutrinos}},
  \href{https://doi.org/10.1088/0954-3899/25/12/302}{\emph{J. Phys.} {\bfseries
  G25} (1999) 2319} [\href{https://arxiv.org/abs/hep-ph/9909302}{{\ttfamily
  hep-ph/9909302}}].

\bibitem{Buras:2012dp}
A.~J. Buras, F.~De~Fazio, J.~Girrbach and M.~V. Carlucci, \emph{{The Anatomy of
  Quark Flavour Observables in 331 Models in the Flavour Precision Era}},
  \href{https://doi.org/10.1007/JHEP02(2013)023}{\emph{JHEP} {\bfseries 02}
  (2013) 023} [\href{https://arxiv.org/abs/1211.1237}{{\ttfamily 1211.1237}}].

\bibitem{Buras:2013dea}
A.~J. Buras, F.~De~Fazio and J.~Girrbach, \emph{{331 models facing new $b \to
  s\mu^+ \mu^-$ data}},
  \href{https://doi.org/10.1007/JHEP02(2014)112}{\emph{JHEP} {\bfseries 02}
  (2014) 112} [\href{https://arxiv.org/abs/1311.6729}{{\ttfamily 1311.6729}}].

\bibitem{Gauld:2013qja}
R.~Gauld, F.~Goertz and U.~Haisch, \emph{{An explicit Z'-boson explanation of
  the $B \to K^* \mu^+\mu^-$ anomaly}},
  \href{https://doi.org/10.1007/JHEP01(2014)069}{\emph{JHEP} {\bfseries 01}
  (2014) 069} [\href{https://arxiv.org/abs/1310.1082}{{\ttfamily 1310.1082}}].

\bibitem{Buras:2015kwd}
A.~J. Buras and F.~De~Fazio, \emph{{$\varepsilon'/\varepsilon$ in 331 Models}},
  \href{https://doi.org/10.1007/JHEP03(2016)010}{\emph{JHEP} {\bfseries 03}
  (2016) 010} [\href{https://arxiv.org/abs/1512.02869}{{\ttfamily
  1512.02869}}].

\bibitem{Dong:2015dxw}
P.~Van~Dong, N.~T.~K. Ngan, T.~D. Tham, L.~D. Thien and N.~T. Thuy,
  \emph{{Phenomenology of the simple 3-3-1 model with inert scalars}},
  \href{https://doi.org/10.1103/PhysRevD.99.095031}{\emph{Phys. Rev.}
  {\bfseries D99} (2019) 095031}
  [\href{https://arxiv.org/abs/1512.09073}{{\ttfamily 1512.09073}}].

\bibitem{Huong:2019vej}
D.~T. Huong, D.~N. Dinh, L.~D. Thien and P.~Van~Dong, \emph{{Dark matter and
  flavor changing in the flipped 3-3-1 model}},
  \href{https://doi.org/10.1007/JHEP08(2019)051}{\emph{JHEP} {\bfseries 08}
  (2019) 051} [\href{https://arxiv.org/abs/1906.05240}{{\ttfamily
  1906.05240}}].

\bibitem{Delbourgo:1972xb}
R.~Delbourgo and A.~Salam, \emph{{The gravitational correction to pcac}},
  \href{https://doi.org/10.1016/0370-2693(72)90825-8}{\emph{Phys. Lett.}
  {\bfseries 40B} (1972) 381}.

\bibitem{AlvarezGaume:1983ig}
L.~Alvarez-Gaume and E.~Witten, \emph{{Gravitational Anomalies}},
  \href{https://doi.org/10.1016/0550-3213(84)90066-X}{\emph{Nucl. Phys.}
  {\bfseries B234} (1984) 269}.

\bibitem{Okada:2016whh}
H.~Okada, N.~Okada, Y.~Orikasa and K.~Yagyu, \emph{{Higgs phenomenology in the
  minimal $SU(3)_L \otimes U(1)_X$ model}},
  \href{https://doi.org/10.1103/PhysRevD.94.015002}{\emph{Phys. Rev.}
  {\bfseries D94} (2016) 015002}
  [\href{https://arxiv.org/abs/1604.01948}{{\ttfamily 1604.01948}}].

\bibitem{Peccei:1977hh}
R.~D. Peccei and H.~R. Quinn, \emph{{CP Conservation in the Presence of
  Instantons}}, \href{https://doi.org/10.1103/PhysRevLett.38.1440}{\emph{Phys.
  Rev. Lett.} {\bfseries 38} (1977) 1440}.

\bibitem{Peccei:1977ur}
R.~D. Peccei and H.~R. Quinn, \emph{{Constraints Imposed by CP Conservation in
  the Presence of Instantons}},
  \href{https://doi.org/10.1103/PhysRevD.16.1791}{\emph{Phys. Rev.} {\bfseries
  D16} (1977) 1791}.

\bibitem{Tanabashi:2018oca}
{\scshape Particle Data Group} collaboration, \emph{{Review of Particle
  Physics}}, \href{https://doi.org/10.1103/PhysRevD.98.030001}{\emph{Phys.
  Rev.} {\bfseries D98} (2018) 030001}.

\bibitem{Ackermann:2012qk}
{\scshape Fermi-LAT} collaboration, \emph{{Fermi LAT Search for Dark Matter in
  Gamma-ray Lines and the Inclusive Photon Spectrum}},
  \href{https://doi.org/10.1103/PhysRevD.86.022002}{\emph{Phys. Rev.}
  {\bfseries D86} (2012) 022002}
  [\href{https://arxiv.org/abs/1205.2739}{{\ttfamily 1205.2739}}].

\bibitem{Aprile:2017iyp}
{\scshape XENON} collaboration, \emph{{First Dark Matter Search Results from
  the XENON1T Experiment}},
  \href{https://doi.org/10.1103/PhysRevLett.119.181301}{\emph{Phys. Rev. Lett.}
  {\bfseries 119} (2017) 181301}
  [\href{https://arxiv.org/abs/1705.06655}{{\ttfamily 1705.06655}}].

\bibitem{Belyaev:2018pqr}
A.~Belyaev, E.~Bertuzzo, C.~Caniu~Barros, O.~Eboli, G.~Grilli Di~Cortona,
  F.~Iocco et~al., \emph{{Interplay of the LHC and non-LHC Dark Matter searches
  in the Effective Field Theory approach}},
  \href{https://doi.org/10.1103/PhysRevD.99.015006}{\emph{Phys. Rev.}
  {\bfseries D99} (2019) 015006}
  [\href{https://arxiv.org/abs/1807.03817}{{\ttfamily 1807.03817}}].

\bibitem{Coleman:1973jx}
S.~R. Coleman and E.~J. Weinberg, \emph{{Radiative Corrections as the Origin of
  Spontaneous Symmetry Breaking}},
  \href{https://doi.org/10.1103/PhysRevD.7.1888}{\emph{Phys. Rev.} {\bfseries
  D7} (1973) 1888}.

\end{thebibliography}\endgroup

\end{document}